\documentclass[journal,twocolumn]{IEEEtran}

\usepackage{cite}

\ifCLASSINFOpdf

\else

\fi

\usepackage{bbm}
\usepackage{amsmath}
\usepackage{amssymb}
\usepackage{outlines}
\usepackage{array}
\usepackage{float}
\usepackage{kotex}
\usepackage{graphicx}
\usepackage{bm}
\usepackage{caption,subcaption}
\usepackage{algorithm}
\usepackage{algpseudocode}
\usepackage{csquotes}
\usepackage{cite}
\usepackage{multicol}
\usepackage{stmaryrd}
\usepackage{tabularx,booktabs}
\usepackage{xcolor}
\newcolumntype{C}{>{\centering\arraybackslash}X} 
\usepackage{lipsum} 

\usepackage{caption,subcaption}

\newsavebox{\measurebox}

\begin{document}
%
\title{Cache-assisted Mobile Edge Computing over Space-Air-Ground Integrated Networks for\\ Extended Reality Applications}
%
%
%

\author{Seonghoon~Yoo,~\IEEEmembership{Student Member,~IEEE,} 

Seongah~Jeong,~\IEEEmembership{Member,~IEEE,}
Jeongbin~Kim,
Joonhyuk~Kang,~\IEEEmembership{Member,~IEEE}
\thanks{Seonghoon Yoo, Jeongbin Kim and Joonhyuk Kang are with the Department of Electrical Engineering, Korea Advanced Institute of Science and Technology, Daejeon 34141, South Korea (e-mail: shyoo902@kaist.ac.kr, kjungbin6560@kaist.ac.kr, jhkang@ee.kaist.ac.kr).}
\thanks{Seongah Jeong is with the School of Electronic and Electrical Engineering, Kyungpook
National University, Daegu 14566, Korea (e-mail: seongah@knu.ac.kr).}}


\maketitle

\begin{abstract}
Extended reality-enabled Internet of Things (XRI) provides the new user experience and the sense of immersion by adding virtual elements to the real world through Internet of Things (IoT) devices and emerging 6G technologies. However, the computational-intensive XRI tasks are challenging for the energy-constrained small-size XRI devices to cope with, and moreover certain data requires centralized computing that needs to be shared among users. To this end, we propose a cache-assisted space-air-ground integrated network mobile edge computing (SAGIN-MEC) system for XRI applications, consisting of two types of edge servers mounted on an unmanned aerial vehicle (UAV) and low Earth orbit (LEO) equipped with cache and the multiple ground XRI devices. For system efficiency, the four different offloading procedures of the XRI data are considered according to the type of information, i.e., shared data and private data, as well as the offloading decision and the caching status. Specifically, the private data can be offloaded to either UAV or LEO, while the offloading decision of the shared data to the LEO can be determined by the caching status. With the aim of maximizing the energy efficiency of the overall system, we jointly optimize UAV trajectory, resource allocation and offloading decisions under latency constraints and UAV's operational limitations by using the alternating optimization (AO)-based method along with Dinkelbach algorithm and successive convex optimization (SCA). Via numerical results, the proposed algorithm is verified to have the superior performance compared to conventional partial optimizations or without cache.
\end{abstract}


\IEEEpeerreviewmaketitle

\section{Introduction}
\IEEEPARstart{E}{xtended reality} (XR) in sixth-generation (6G) networks are envisioned to support numerous Internet of Things (IoT) applications, including autonomous driving, remote surgery and military operations, which are leveraged by advances in wireless communications \cite{6G, 6G_IoT}. XR is an innovative technique that includes virtual reality (VR), augmented reality (AR) and mixed reality (MR), referring to various experiences that combine digital and real-world aspects. Recently, due to the evolved cellular communication technologies, the smart IoT devices can communicate with high-traffic XR data in real-time, and the concept of an IoT-based XR system or an XR-based IoT system is called as XR-enabled IoT (XRI) \cite{IoT_XR_GEM, IoT_XR_VRW, ar_enabled_IoT, IoT_XR_IoTJ}. The potential use-case scenarios of XRI systems are of interest to both industry and academic studies since they can improve human-to-thing and human-to-human relationships, enabling the design of future personal and industrial applications \cite{IoT_XR_IoTJ, survey_5G_iot}. However, the XR-based applications are, in general, computationally intensive, and therefore computing on the battery-limited small-size IoT device itself cannot satisfy the stringent Quality of Services (QoS). To this end, mobile edge computing (MEC) is being investigated so as to minimize the computing time by offloading to the nearby edge servers\cite{iot_mec}. Interestingly, compared to the typical offloaded tasks in MEC systems, the XRI data has collaborative properties, which consist of shared data and private data \cite{osvaldo, globecom_xr, xr_mec, powercontrol_ar}. For example, the XR tasks required for image obtaining, rendering, and tracking can be individually processed in a general manner. Meanwhile, Mappers and Object recognizers related to environmental and background processing are required to collect inputs from all devices in the same region and to broadcast them to all users \cite{collabo_xr}. Therefore, the common information such as the virtual and physical backgrounds or the object observed by terrestrial XRI devices needs to be shared, while the individual information such as rendering and video processing is processed by each XRI device, ensuring privacy.

For providing high-throughput and ultra-low-latency services, non-terrestrial infrastructures such as aerospace networks with low-Earth-orbit (LEO) satellites and unmanned aerial vehicles (UAVs) have been receiving increased attention in MEC systems \cite{sooyeob, sagin_mec1,ADMM,  sagin_mec11, sagin_mec2}. The UAVs can easily secure a highly probable line-of-sight (LoS) of wireless channels \cite{Jeong2018MEC} by controlling their trajectory. On the other hand, LEO satellites offer seamless global coverage at much higher altitudes, for which global companies such as Telesat, SpaceX Starlink, OneWeb Satellites and Amazon Kuiper \cite{sagin_company} have recently invested by launching their own countless LEOs. With the real-time offloading opportunities within global coverage based on the non-terrestrial infrastructures, the XRI services have the potential to become a reality in the foreseeable future \cite{ADMM, Antennagain}. 

Moreover, the performance of MEC systems can be further enhanced by adopting a cache \cite{cache_highref2, cache_highref} that can reduce the unnecessary data transfer and task computation, which is crucial for XRI applications to require ultra-low responding time. Also, this is more emphasized in the non-terrestrial networks that are highly sensitive to offloading latency due to the communication over the space-air-ground links spanning more than 600 km \cite{satellite_cache, leo_cache}. In that sense, the caching technique is very attractive for addressing insufficient computing and transmission processes by storing in-process resources for later use \cite{offload_cache}. For the XRI applications with common shared data among service users, the concept of edge caching is promising since similar types of video files are frequently accessed. For instance, the VR video content to be transmitted in the form of temporally and spatially small VC (Video Chunk) can be divided and stored as the content in the form of small data units called monoscopic VC (MVC) according to the popularity of the VC, which reduces the significant amount of operating time \cite{vr_caching}. In this kind of caching strategies, the optimization of caching placement is a key issue since the cache has a limited capacity, which makes the edge server decide which files to store.

\subsection{Related Works}
\subsubsection{MEC for XRI applications}
A recent line of work about the \textit{MEC} systems for XRI applications has primarily focused on reducing energy consumption or total latency. Specifically, the authors in \cite{osvaldo, globecom_xr} aim to minimize mobile energy consumption under delay constraint taking into account XR's collaborative properties. In \cite{osvaldo}, it is noted that the users of AR applications have computational tasks that need to be shared with other users and simultaneously that need to be separated. The authors in \cite{globecom_xr} propose a multi-user fog computing system, allowing mobile users running applications with shared data to choose between partial offloading to cloudlets and local execution. In \cite{xr_mec}, an architecture supporting multiple XR applications for mobile subscribers has been introduced, encouraging device cooperation to further reduce latency and energy consumption. In \cite{powercontrol_ar}, the energy consumption reduction of MEC systems for XR applications is focused on optimizing the transmit power and frame resolution of objects. With the rapid development of non-terrestrial networks (NTNs), the \textit{SAGIN-MEC} systems have been actively studied \cite{sooyeob, sagin_mec1,ADMM, sagin_mec11, sagin_mec2} to address the computational-intensive tasks or delay-sensitive tasks. In \cite{sooyeob}, the authors propose a SAGIN network considering LEO and UAV edge computing in a marine IoT system to jointly optimize bit allocation and UAV trajectory planning to minimize UAV's energy consumption. In \cite{sagin_mec1}, the satellites equipped with cloud servers and UAVs equipped with edge servers are considered for marine IoT systems. In order to minimize the maximum computational delay among IoT devices, the association control and the UAV's resources and deployment are jointly optimized. The authors in \cite{Antennagain} propose a SAGIN-MEC system that minimizes the total energy cost of UAVs by processing the delay-sensitive IoT data collected from UAV and processing the rest by LEO. 
\subsubsection{Caching for XRI applications}
Recently, the \textit{caching} techniques have been developed for the edge server in various ways. In particular, several studies have been introduced for XRI applications to optimize the content placement or caching policy \cite{cache_flow, vr_caching, leo_cache, offload_cache, vr_transcoding_cache, chen2022joint}. In \cite{leo_cache}, the system
model that jointly optimizes resource allocation, UAV trajectory and cache placement in content delivery networks is proposed to maximize the minimum achievable throughput per user. Assuming the placement-then-delivery method, the binary content placement variables are optimized under the limited cache storage size. The authors in \cite{offload_cache} propose a collaborative caching and offloading system for compute-intensive and latency-sensitive tasks from the edge cloud under computing and storage resource constraints to minimize energy costs on mobile devices. In \cite{vr_caching}, the scalable content placement scheme is designed by considering both caching and computing resources to obtain the optimal caching solution for high-quality VR video streaming. The authors in \cite{cache_flow} propose a cache-assisted collaborative task offloading and resource allocation strategy by meta-reinforcement learning. Also, the transcoding-enabled caching framework for VR video in the edge network is developed in \cite{vr_transcoding_cache} to minimize the average latency. In \cite{chen2022joint}, the authors investigate the caching and computing resource management system, while satisfying the latency requirements for processing the spatio-temporal varying and location-aware services, which are the major characteristics of XRI applications. 

As outlined above, most SAGIN-MEC systems consider general tasks rather than specific services, the performance of which can be further optimized by utilizing the characteristics of tasks for offloading procedure. In addition, the existing SAGIN-MEC studies consider either LEO or UAV-mounted cloudlets. Especially, since the SAGIN-MEC systems for XRI applications can be expected to be useful for the emergency and disaster responses in military, medical or industrial services, where the seamless services are essential, the aerial and spatial cloudlets need to be efficiently and adaptively performed. Consequently, the usage of caching can provide the considerable impact in SAGIN-MEC systems for XRI applications, which has not been discussed so far. 

\subsection{Main Contributions}
In this paper, we propose the cache-assisted SAGIN-MEC systems for XRI applications, as illustrated in Fig. \ref{System model}, where both LEO equipped with the cache and UAV are mounted with cloudlets. In the proposed set-up, we consider the four types of offloading and caching procedures considering the collaborative properties of XRI applications and the offloading task requirements. In the offloading procedure, the delay-sensitive tasks, for instance, can be offloaded to the UAV with the relatively short inter-node distance, while the computation-intensive tasks can be offloaded to LEO with the sufficient computing capability. Here, the offloaded shared tasks can be stored via caching at the LEO, which reduces the redundant computing and communication process. The remaining shared data, i.e., cache-miss data, is offloaded to the LEO with cloudlet that has sufficient computing capability. For offloading the private data, which is handled individually, either UAV or LEO with cloudlet needs to be optimally selected based on the system's objective. In this work, we aim to maximize the system energy efficiency by jointly optimizing the UAV trajectory, resource allocation and offloading decision. To tackle with the non-convexity of the formulated problem, the alternating optimization (AO)-based method is proposed along with efficient algorithms such as Dinkelbach algorithm \cite{dinkelbach, FP_SCA} and successive convex optimization (SCA) \cite{Jeong2018MEC, RuiEE}. The main contributions of this paper are summarized as follows.
\begin{outline}
    \1 The cache-assisted MEC systems with LEO and UAV-mounted cloudlets are proposed based on the collaborative characteristics of XRI data. In particular, according to the offloading decision and caching results, the four types of offloading procedures are introduced such as i) Cache Hit and offloading Priviate Data to UAV (CH-PD2U), ii) Cache Hit and offloading Private Data to LEO (CH-PD2L), iii) Cache Miss and offloading Private Data to UAV (CM-PD2U) and iv) Cache Miss and offloading Private Data to LEO (CM-PD2L). 
    \1 We formulate the optimization problem to maximize the energy efficiency of the overall system by jointly optimizing the UAV's trajectory, resource allocation and offloading decision, which encompasses the four offloading scenarios of the cache-assisted SAGIN-MEC systems. The proposed algorithm to achieve the optimal solution is developed based on AO method \cite{Zhang2018AO} coupled with Dindelbach \cite{dinkelbach, FP_SCA} and SCA \cite{Jeong2018MEC, RuiEE} approaches, whose convergence and complexity are analytically and numerically discussed.
    \1 Via simulations, the proposed joint optimal solution can achieve about 35\% more energy efficiency on average compared to the partial optimization schemes. Moreover, the use of cache can provide the 15\% performance improvement compared to the case without the cache.
\end{outline}

The remainder of this paper is organized as follows. Section II presents the system model including the caching and offloading model of the SAGIN-MEC system for XRI applications. In Section III, we formulate and tackle the optimization problem to maximize the energy efficiency of the overall system. Section IV presents the numerical results, and conclusions are provided in Section V.








\section{System Model}
\subsection{System Model Set-up}
 \begin{figure}[t] 
\begin{center}
\includegraphics[width=1\columnwidth]{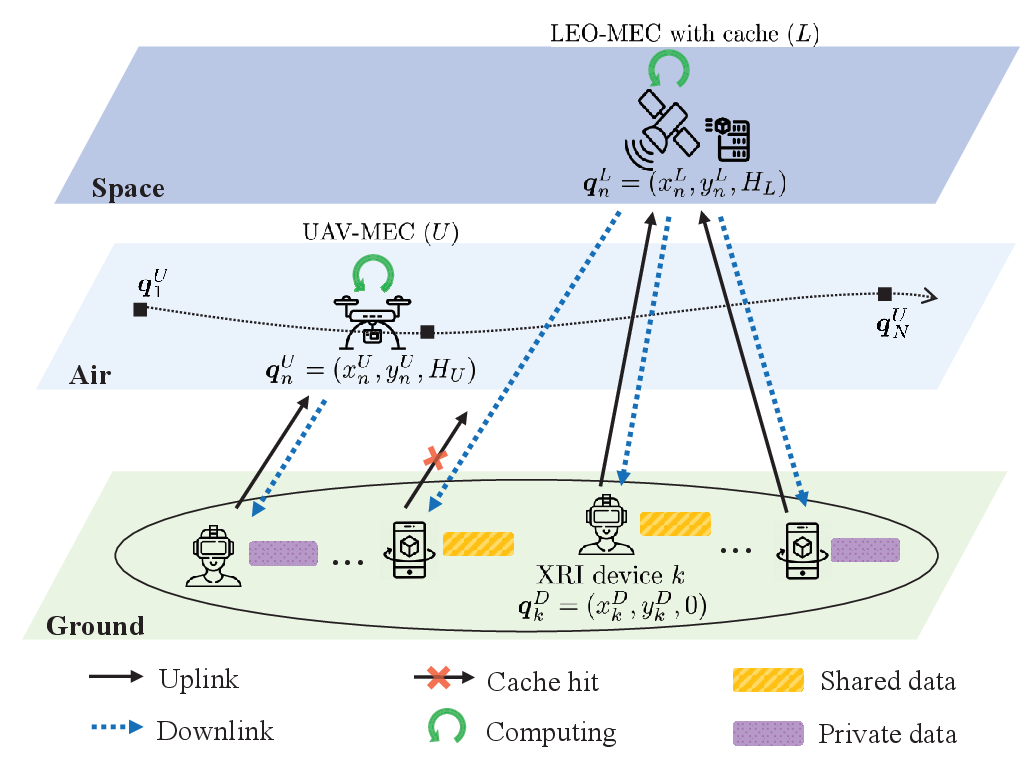}
\caption{System model of the cache-assisted SAGIN-MEC systems for XRI applications.}
\label{System model}
\end{center}
\end{figure}
We consider a cache-assisted SAGIN-MEC system for XRI applications. The XRI devices have, in general, the traditional image processing to be computed, which is composed of Video source, Renderer, Object recognizer, Trackers and Mappers \cite{osvaldo}. Specifically, Video source obtains the raw image frame from the XRI device's camera, and Renderer prepares the processed frames to display the processed frame, both of which can be computed locally. On the other hand, Object recognizers recognize objects in the environment, Trackers track information that users move around in the environment, and Mappers build models in the environment. They have computationally intensive tasks and therefore are subject to be offloaded. The examples of XRI services can be mainly thought of as medical service support, 3D design, gaming, etc \cite{van2010survey}, which have shared inputs and outputs and pertain to Tracker, Mapper and Object recognizer. By following \cite{osvaldo}, in this paper, two types of data are considered such as shared data and private data, the latter of which is personal data. The shared data is first offloaded, its calculation is finished at the server, and it can be distributed to local users at once in the form of a multicast manner via downlink. The processing of private data is performed sequentially after processing the shared data, and its computing results are delivered to the corresponding users in a unicast manner.

As shown in Fig. \ref{System model}, in order to provide the XRI application services within global communication and computation coverage, the system of interests has three layers, consisting of space area with LEO satellite-mounted servers, aerial area with UAV-mounted servers and ground area with multiple XRI devices.  
The LEO satellite with the large coverage area and the high computational capacity is employed to process both shared and private input data globally at once, while the UAV supports the layer between XRI devices and LEO satellite to process only private input data locally. In the private data offloading procedure, the XRI devices need to decide the computing server located at either the LEO satellite or UAV. Note that each XRI device transmits both shared data and private data of the file during mission time.   
Moreover, the LEO satellite is assumed to be equipped with a cache to store and deliver the results of the offloaded shared data, e.g., the static environment data, to the ground XRI devices within the entire coverage area, whenever they are requested, which can reduce the unnecessary repeated computing.  

\begin{figure}[t] 
\begin{center}
\includegraphics[width=1\columnwidth]{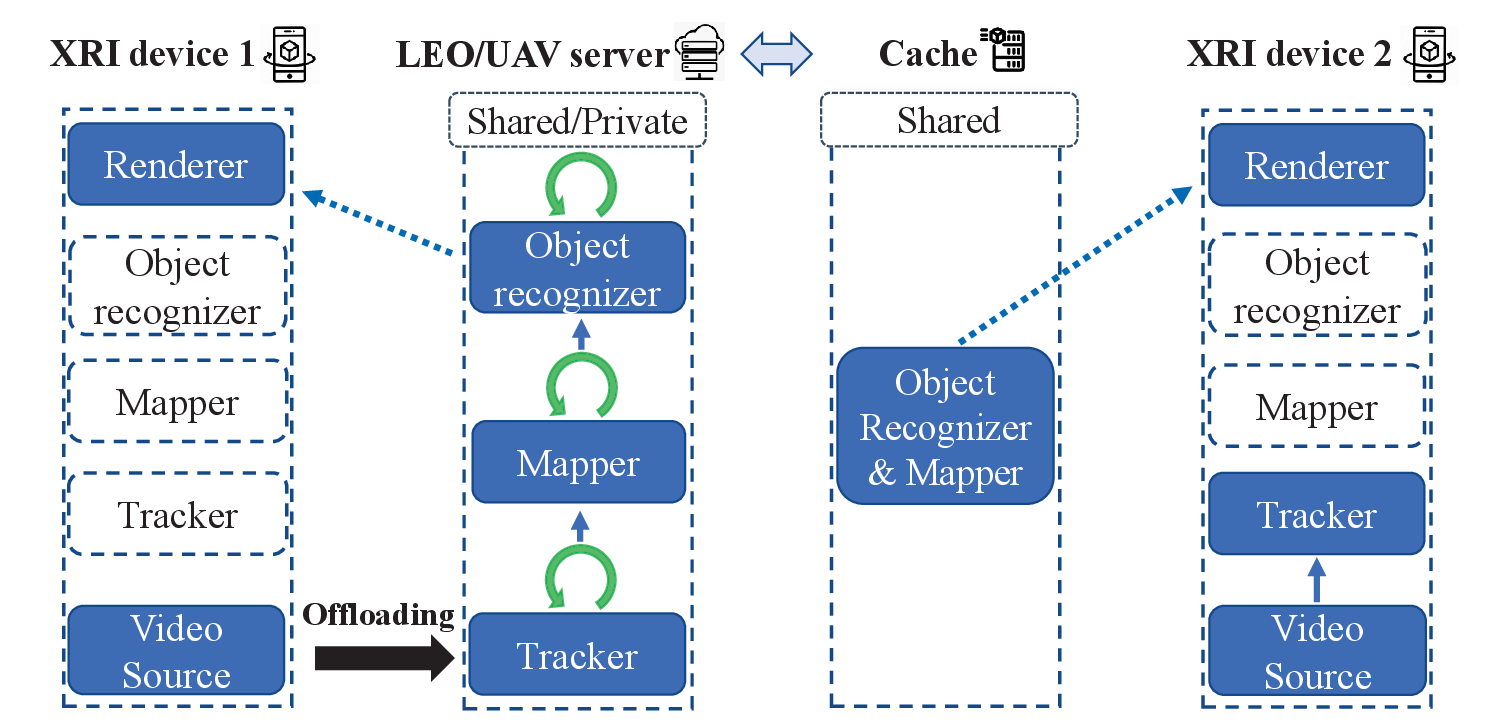}
\caption{Example of a component flow.}
\label{XR_SAGIN_cache}
\end{center}
\vspace{-15pt}
\end{figure}

Fig. \ref{XR_SAGIN_cache} describes the example of the component flow of the SAGIN-MEC system for XRI applications. The components to require the intensive computation such as Tracker, Mapper and Object recognizer perform the offloading procedure to the edge server. The computation of private data is carried out at LEO or UAV, while the shared data of Mapper and Object recognizer require centralized computing, which is offloaded to the LEO server for its computation, or is checked via cache about whether the computation results exist or not. The computation outputs resulting from either edge server or cache are sent to the XRI device as requested. For simplicity, all the nodes are assumed to be equipped with a single antenna, and a single pair of LEO satellite and UAV is adopted for the offloading process of $K$ XRI devices. We assume that all XRI devices within coverage use Time Division Duplex (TDD) over frequency-flat fading channels. In the following, we use the notations of $D$, $L$ and $U$ for XRI devices, LEO and UAV, respectively.  
 \begin{figure*}[t] 
\begin{center}
\includegraphics[width=1.8\columnwidth]{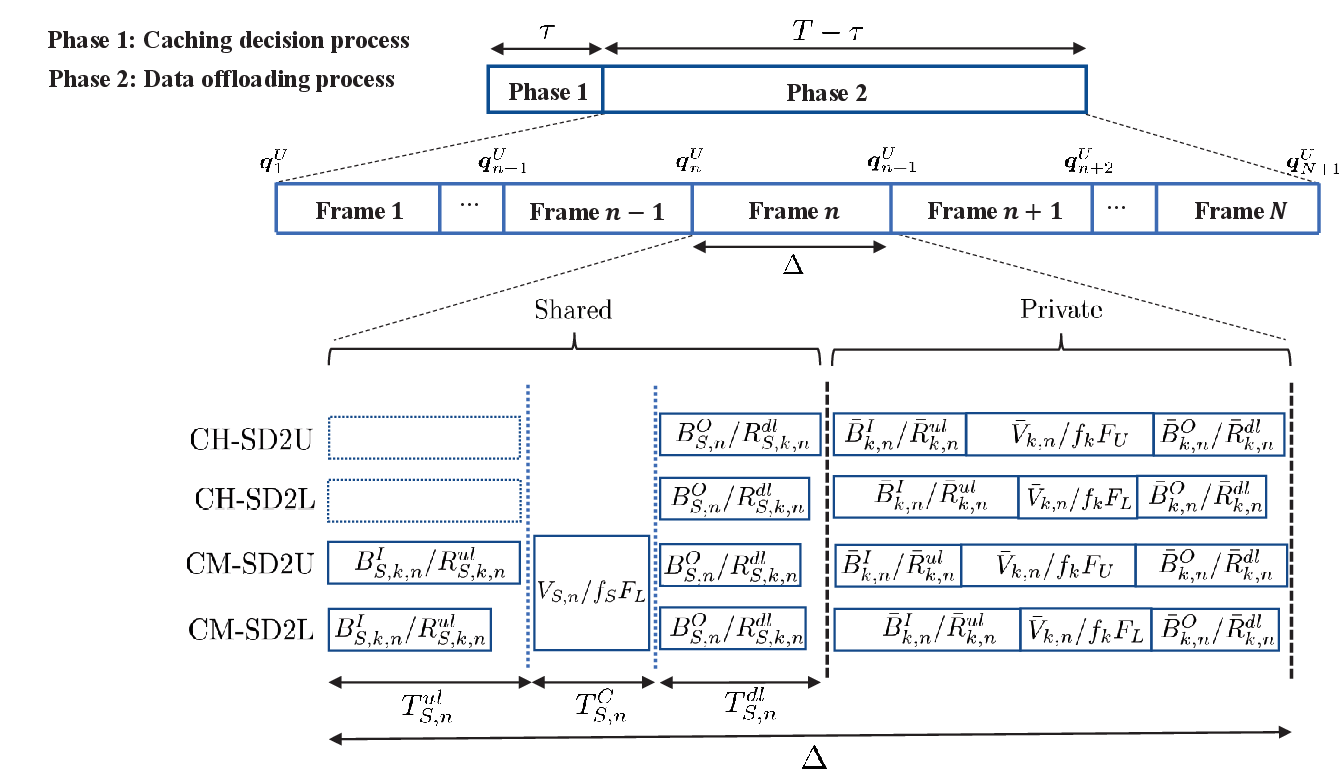}
\caption{Frame structure with Phase 1 for caching decision process and Phase 2 for data offloading process. According to the caching result and the offloading decision, four types of scenarios are considered: i) Cache Hit and offload Private Data to UAV (CH-PD2U), ii) Cache Hit and offload Private Data to LEO (CH-PD2L), iii) Cache Miss and offload Private Data to UAV (CM-PD2U), iv) Cache Miss and offload Private Data to LEO (CM-PD2L).}
\label{time_frame}
\end{center}
\end{figure*}
\subsection{Frame Structure}
In Fig. \ref{time_frame}, we consider the discrete-time frame model, consisting of two phases according to the caching process. Phase 1 determines whether the file $f \in \{1,2,..., F\}$ to be offloaded by XRI device $k \in \mathcal{K}\triangleq \{1,2,...,K\}$ is cached in LEO, where $F$ is the number of file types handled by XRI devices. In this phase, the data transmitted in both uplink and downlink is the indicator data, which is assumed to be processed within $\tau$, with $\tau$ being the total times of Phase 1. In Phase 2, the offloading decision on whether or not to utilize the intermediate layer with UAV server is optimized based on the cache hit result obtained in Phase 1, and then each XRI device offloads the shared data and the private data sequentially. The detailed flow of Phase 1 and Phase 2 is further explained in Section II-C. The mission time $T-\tau$ of Phase 2 is divided into $N$ time frames, each of which has equal duration $\Delta$ to satisfy the condition of $T-\tau=N\Delta$. 
In Phase 2, the orthogonal multiple access, e.g., OFDMA, is assumed, for which, the bandwidth $W^{ul}$ and $W^{dl}$ of uplink and downlink are equally divided to $K$ XRI devices with no interference. Assuming $\tau \ll T-\tau$ in Phase 1, which can be ignored, we focus on Phase 2 for the optimization of resource allocation and offloading decisions.

For Phase 2, we consider the following procedure. XRI device $k$ transmits input bits to the flying edge server at either LEO or UAV. For example, the input bits to Object recognizer of the server can be sent by any of the XRI devices in the same coverage. Each XRI device $k$ transmits $B_{S,k,n}^I$ bits of shared data at the $n$-th frame as well as $\bar{B}_{k,n}^I$ bits of private data that need to be uploaded exclusively by XRI device $k$, where the set of $N$ time frames is represented as $\mathcal{N} \triangleq \{1,2,...,N\}$. These edge servers of LEO and UAV execute the computation with a certain cycle regardless of the time frame $n$, and the computational effort of the edge server is spent on producing the output bits of interest to all devices. We assume that $V_{S,n}=\varepsilon_S\times \sum_{k \in \mathcal{K}}(1-c_k)B_{S,k,n}^I$ CPU cycles for offloaded shared data, where $c_k$ indicates whether to be cached or not at LEO satellite. Whereas, $\bar{V}_{k,n}=\bar{\varepsilon} \times \bar{B}_{k,n}^I$ CPU cycles are to be executed for each XRI device $k$ \cite{osvaldo}, where $\varepsilon_S$ and $\bar{\varepsilon}$ are parameters of CPU cycles per bit of shared input data and private input data, respectively. Lastly, LEO or UAV server transmits the output bits to the XRI device $k$ in the downlink, e.g., the XRI devices in the same region may need the output data from Mapper to update a map. We assume $B_{S,n}^O$ output bits can be transmitted in multicast manner to all XRI devices, while $\bar{B}_{k,n}^O$ bits need to be transmitted to each XRI device in a unicast manner, where $B_{S,n}^O$ and $\bar{B}_{k,n}^O$ denote bits of shared output data and private output data, respectively.
 \begin{figure*}[t] 
\begin{center}
\includegraphics[width=2.1\columnwidth]{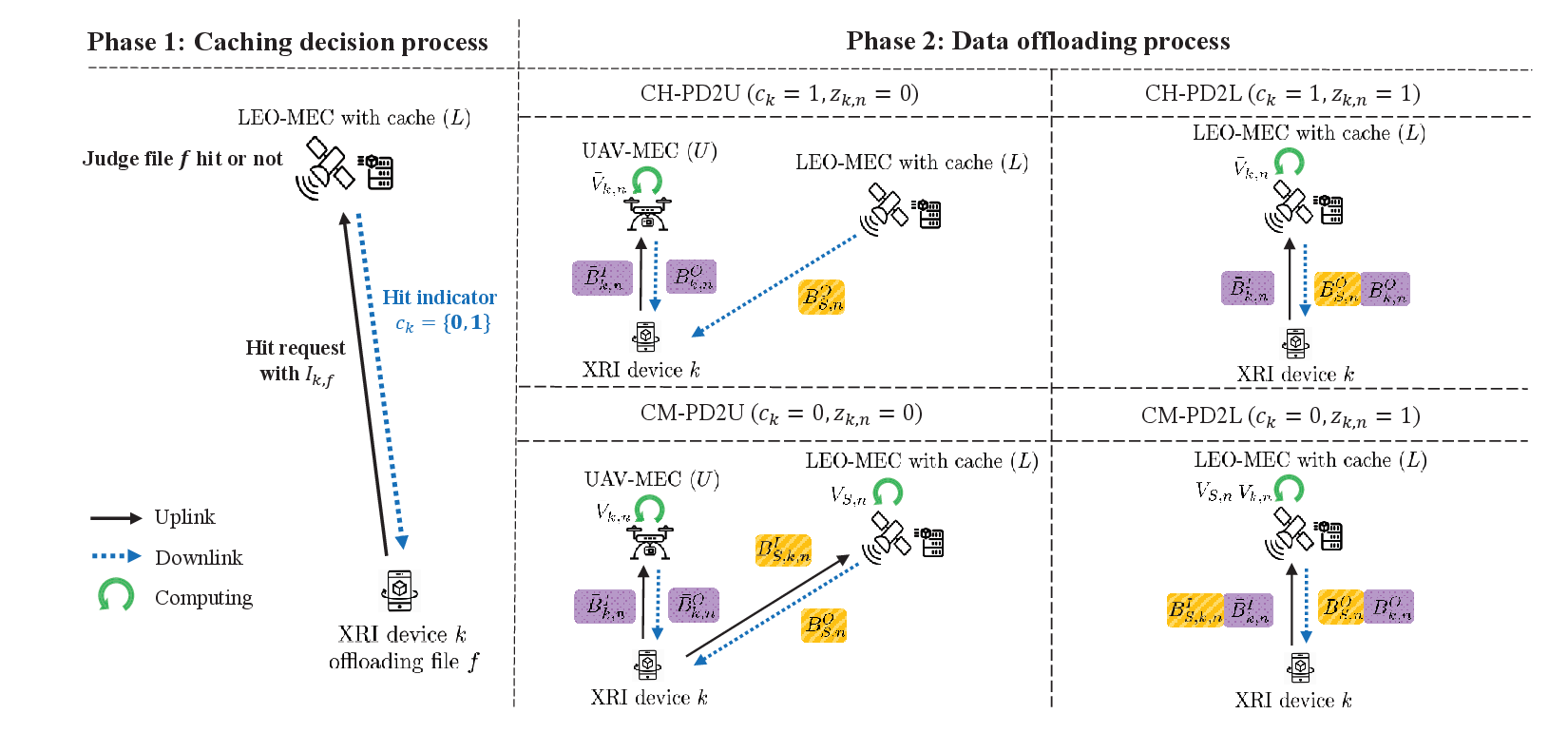}
\caption{Caching and offloading model for SAGIN-MEC system.}
\label{offload_cache}
\end{center}
\end{figure*}
\subsection{Aerospace Trajectory Model}
In real applications, the LEO satellite flies along with the fixed orbit trajectory $\boldsymbol{q}^L_n$, and has a geometric relationship with the ground XRI devices \cite{ADMM}. In particular, the LEO moves with a constant orbital speed $v_L$ along the arc with the length of $l= 2(R_E+H_L)$ to communicate with both XRI devices and UAV, where $R_E$ is the radius of Earth and $H_L$ is the fixed altitude of LEO satellite. The trajectory $\boldsymbol{q}^L_n$ of LEO satellite is expressed as $\boldsymbol{q}^L_n=(x_n^L, y_n^L, H_L$), and the time limit $T_v$ for coverage can be calculated as
\begin{equation}
    T_v=\frac{l}{v_L}=\frac{2(R_E+H_L)\phi}{v_L},
\end{equation}
where $\phi$ is the angle of the satellite coverage and $v_L$ is the speed of the LEO satellite. We assume that the LEO is always available for seamless services during the overall mission time $T$, i.e., $T_v \geq T$.
 The UAV proceeds the offloading procedure at the relatively low altitudes $H_U$ to compensate for the delay problem of LEO-based offloading due to the long distance of LEO communication with $H_L \gg H_U$. The trajectory $\boldsymbol{q}_n^U$ of UAV is expressed as $\boldsymbol{q}_n^U=(x_n^U,y_n^U,H_U)$, for $n \in \mathcal{N}$. For the mission of XRI applications, the initial location $\boldsymbol{q}_1^U$ and the final location $\boldsymbol{q}_N^U$ of the UAV $U$ are assumed to be predetermined. Following the discrete state-space model in \cite{Hybrid,Jeong2018MEC} the velocity and trajectory variables of UAV can be defined as
 \begin{equation}
 \label{vel}
     \boldsymbol{v}_{n+1}^U = \boldsymbol{v}_n^U+\boldsymbol{a}_n^U\Delta 
 \end{equation}
 and
  \begin{equation}
   \label{traj}
     \boldsymbol{q}_{n+1}^U = \boldsymbol{q}_n^U+\boldsymbol{v}_n^U\Delta+\frac{1}{2}\boldsymbol{a}_n^U\Delta^2,
 \end{equation}
for $n \in \mathcal{N}\cup\{0\}$, where $\boldsymbol{a}_n^U$ and $\boldsymbol{v}_n^U$ are accelaration and velocity vector of UAV, respectively. Also, its maximum speed and acceleration constraint is given as
\begin{equation}
\label{vel_max}
\|\boldsymbol{v}_n^U\|=\frac{\|\boldsymbol{q}_{n+1}^U-\boldsymbol{q}_n^U\|}{\Delta}\leq v_{\max}
\end{equation}
and
\begin{equation}
\label{acc_max}
\|\boldsymbol{a}_n^U\|=\frac{\|\boldsymbol{v}_{n+1}^U-\boldsymbol{v}_n^U\|}{\Delta}\leq a_{\max},
\end{equation}
where $v_{\max}$ and $a_{\max}$ are the maximum velocity and acceleration constraint, respectively. The XRI device $k$ is assumed to be fixed at $\boldsymbol{q}^D_k=(x^D_{k},y^D_{k},0)$ during the entire mission time, for $k \in \mathcal{K}$.

\subsection{Channel Model}
By following \cite{rician_12}, Rician fading is adopted for the ground-to-air (G2A) channel, and the channel power gain between XRI device $k$ and UAV $U$ in frame $n$ can be written as 
\begin{equation}
        h_{k,n}^{D,U}(\boldsymbol{q}^U_n)=\dfrac{\beta_0}{\lVert \boldsymbol{q}_n^U-\boldsymbol{q}_n^D\rVert^2+{H_U}^2}\chi^\textrm{G2A},
\end{equation}
for $\forall k \in \mathcal{K}$, where $\beta_0$ denotes the received power at the reference distance $d_0=1$ m of the G2A link. Also, $\chi^\textrm{G2A}$ is a small scale fading component in the G2A environment with Rician factor $K^{\textrm{G2A}}$ defined as $\chi^\textrm{G2A}=\sqrt{{K^{\textrm{G2A}}}/{(K^{\textrm{G2A}}+1)}}\gamma+\sqrt{{1}/{(K^{\textrm{G2A}}+1)}}\Tilde{\gamma}$ \cite{rician_12}, \cite{rician_eq}, where $\gamma$ denotes the deterministic Line-of-sight (LoS) component with $|\gamma|=1$, and $\Tilde{\gamma}$ 
is a circularly symmetric complex Gaussian (CSCG) random variable for non LoS (NLoS) components.

Since we assume XRI devices on the ground can communicate directly with the satellites \cite{G2S_MIMO2016}, and thus all the distance information between XRI devices and LEO satellite is assumed to be known, we can define channel power gain $h_{k,n}^{D,L}$ of ground-to-space (G2S) link as
\begin{equation}
  h_{k,n}^{D,L}=\dfrac{G\beta_1}{\lVert \boldsymbol{q}_n^D-\boldsymbol{q}_n^L\rVert^2+{H_L}^2}\chi^\textrm{G2S},
\end{equation}
where $\beta_1$ denotes the reference channel power gain of the G2S link, $\chi^\textrm{G2S}$ is small-scale fading component with Rician factor $K^{\textrm{G2S}}$, and $G$ is an antenna gain for the long-distance communication by the XRI device or LEO satellite, whose value can be freely adjusted according to the antenna configuration \cite{sooyeob, Antennagain}.

\subsection{Caching and Offloading Model}
In this section, we describe the caching and offloading model for the cache-assisted SAGIN-MEC networks as illustrated in Fig. \ref{offload_cache}.

For Phase 1 of the caching decision process, each XRI device $k$ requests cache decision to LEO with the data bits $I_{k,f}$ that contain the information about XRI device $k$ to process file $f$ over the uplink. Then, the LEO satellite determines whether the file $f$ is cached or not. The way of cache hit procedure is supposed to be performed to check whether the received cache file index is in the cache storage by following \cite{caching_decision}. Specifically, $c_k=1$ denotes the case, where the task to be processed sent by the XRI device is stored at the cache of LEO, otherwise $c_k=0$. By adopting the content popularity model \cite{satellite_cache, caching_decision}, which considers the Zipf distribution, we determine the caching criteria. Accordingly, the probability for content $f$ being requested by XRI device $k$ is defined as \begin{equation}
P_{k,f}=\dfrac{f^{-\varrho}}{\sum_{i=1}^Fi^{-\varrho}},
\end{equation}where $\varrho$ is the Zipf skewness factor. Considering the demanded probability and cache storage, the LEO determines the caching decision $c_k$, for $k \in \mathcal{K}$. After that, the LEO executes all files that decided to be stored in the cache for a sufficient amount of time. As soon as the computing process ends, LEO retransmits a one-bit output data $c_k$ to each XRI device $k$ by unicast manner. The XRI device determines whether to offload the shared data in Phase 2 based on the received caching indicator result $c_k$.

For Phase 2 of the data offloading process, the decision needs to be made on whether to offload both shared data and private data in frame $n$, according to the choice of $c_k$ and $z_{k,n}$, where $z_{k,n}=1$ if IoT $k$ offloads the private data to the LEO, while $z_{k,n}=0$ if IoT $k$ offloads the private data to the UAV, for $\forall k \in \mathcal{K}$ and $\forall n \in \mathcal{N}$. When offloading private tasks, the procedure is organized so that latency-sensitive tasks can be offloaded to UAV with relatively low transmission latency, and computation-intensive tasks can be offloaded to LEO with sufficient computing capability. The possible four types of flows are illustrated in Fig. \ref{offload_cache}. Due to the operational and physical constraints such as the locations of UAV and LEO as well as the limited operation capability of UAV, the XRI device $k$ needs to select either LEO or UAV so as to offload the private data. In the following, we provide the detailed description for the four different scenarios.

\begin{itemize}
    \item CH-PD2U with $c_k=1$ and $z_{k,n}=0$: In the CH-PD2U case, since the LEO has the hit output data of XRI device $k$, after waiting for cache miss data to be executed, the corresponding result $B_{S,n}^O$ bits are transmitted in the multicast manner via downlink. And then, the private data $\bar{B}_{k,n}^I$ bits are offloaded to the UAV, each data operation is performed with $\bar{V}_{k,n}$ CPU cycles, and the XRI device $k$ receives $\bar{B}_{k,n}^O$ bits via downlink within the allotted time.

    \item CH-PD2L with $c_k=1$ and $z_{k,n}=1$: In the case of the shared data, the process is performed equivalently as CH-PD2U, while the private data of $\bar{B}_{k,n}^I$ bits are offloaded to LEO for computing.

    \item CM-PD2U with $c_k=0$ and $z_{k,n}=0$: In the CM-PD2U case, since the shared data to be processed in Phase 1 is not stored at LEO cache, the XRI device $k$ first offloads the shared data $B^I_{S,k,n}$ bits to LEO and the LEO computes it with $V_{S,n}$ CPU cycles and transmits the $B^O_{S,n}$ bits in multicast manner. And then, the offloading procedure of the private data $\bar{B}_{k,n}^I$ bits is performed in the same way with that of CH-PD2U.

    \item CM-PD2L with $c_k=0$ and $z_{k,n}=1$: In the case of the shared data, the process is performed equivalently as CM-SD2U, while the private data is processed in the same way of CH-PD2L.
\end{itemize}

We can calculate the uplink data rate between XRI device $k$ and edge server for the shared data transmission as
\begin{equation}
    {R}_{S,k,n}^{ul}=\frac{W^{ul}}{K}\log_2 \bigg( 1+\frac{p_Dh_{k,n}^{D,L}}{N_0W^{ul}/K}\bigg),  \forall n \in \mathcal{N}, \forall k \in \mathcal{K}.
\end{equation}
Similarly, the downlink data rate for XRI device $k$ in a multicast manner for the transmission of the shared output data is written as
\begin{equation}
    R_{S,k,n}^{dl}=W^{dl}\log_2 \bigg( 1+\frac{p_Lh_{k,n}^{D,L}}{N_0W^{dl}}\bigg), \forall n \in \mathcal{N}, \forall k \in \mathcal{K}.
\end{equation}
For the separate data transmission, we can calculate the uplink and downlink data rate in a unicast manner as 
\begin{equation}
\begin{aligned}
    & \bar{R}_{k,n}^{ul}(\boldsymbol{q}_{n}^{U},z_{k,n})=\\&\frac{W^{ul}}{K}\log_2 \bigg( 1+\frac{p_D\big(z_{k,n}h_{k,n}^{D,L}+(1-z_{k,n})h_{k,n}^{D,U}(\boldsymbol{q}_{n}^{U})\big)}{N_0W^{ul}/K}\bigg),
\end{aligned}
\end{equation}
and
\begin{equation}
\begin{aligned}
    & \bar{R}_{k,n}^{dl}(\boldsymbol{q}_{n}^{U},z_{k,n})=\\&\frac{W^{dl}}{K}\log_2 \bigg( 1+\frac{\big(z_{k,n}p_Lh_{k,n}^{D,L}+(1-z_{k,n})p_Uh_{k,n}^{D,U}(\boldsymbol{q}_{n}^{U})\big)}{N_0W^{dl}/K}\bigg),
\end{aligned}
\end{equation}
for $\forall n \in \mathcal{N}$ and $\forall k \in \mathcal{K}$, where $N_0$ denotes the available power spectral density, and the transmit powers of XRI devices, UAV and LEO satellite are defined as $p_D$, $p_U$ and $p_L$, respectively.

As seen in the frame structure of Fig. \ref{time_frame}, the procedure for communication and computation of the shared data is carried out first, followed by the offloading procedure of the private tasks. For the shared data, the initial transmission time are needed as 
\begin{equation}
T_{S,n}^{ul}=\max_k (1-c_k)B^I_{S,k,n}/R^{ul}_{S,k,n},    
\end{equation}
and the execution time for the shared of LEO satellite is  
\begin{equation}
T_{S,n}^C=V_{S,n}/(f_SF_L),  
\end{equation}
where $F_L$ is the processing power of LEO. The initial downlink transmission time to multicast $B_{S,n}^O$ bits can be computed as
\begin{equation}
T_{S,n}^{dl}=B_{S,n}^O/\min_kR_{S,k,n}^{dl}.
\end{equation}
For the communication and computation of private data, the transmission time of $\bar{B}_{k,n}^I$ bits is $\bar{B}_{k,n}^I/ \bar{R}_{k,n}^{ul}(\boldsymbol{q}_{n}^{U},z_{k,n})$, and the time needed to execute with $\bar{V}_{k,n}$ CPU cycles is $\bar{V}_{k,n}/(f_kF_U)$ or $\bar{V}_{k,n}/(f_kF_L)$ with the processing power of $F_U, F_L$ according to the types of edge server target, i.e., UAV or LEO. The downlink transmission time to unicast $\bar{B}_{k,n}^O$ bits can accordingly be computed as $\bar{B}_{k,n}^O/\bar{R}_{k,n}^{dl}(\boldsymbol{q}_{n}^{U},z_{k,n})$.

\section{Problem Formulation and Proposed Algorithm}
\subsection{Problem Formulation}
The aim of this paper is to maximize the energy efficiency of the proposed cache-assisted SAGIN-MEC system, which is defined as the sum of the total communication bits normalized by the UAV propulsion energy consumed at each frame, by jointly optimizing offloading decisions $\{z_{k,n}\}_{k\in \mathcal{K},n\in \mathcal{N}}$, the shared data input bit $\{B_{S,k,n}^I\}_{k \in \mathcal{K},n \in \mathcal{N}}$, the UAV trajectory $\{\boldsymbol{q}_{n}^U,\boldsymbol{v}_{n}^U, \boldsymbol{a}_{n}^U\}_{n \in \mathcal{N}}$ and the shared data transmission latency $\{\boldsymbol{T}_n\}\triangleq\{T_{S,n}^{ul}, T_{S,n}^{dl}\}_{n\in \mathcal{N}}$. To this end, we formulate the problem as follows:

\begin{subequations}\label{p1}
\begin{eqnarray}
&&\hspace{-1.3cm}\textrm{(P1)}: \hspace{-0.2cm}\max_{\substack{z_{k,n}, B_{S,k,n}^I,\\\boldsymbol{q}_n^U,\boldsymbol{v}_n^U,\boldsymbol{a}_n^U,\boldsymbol{T}_n}} \sum_{n \in \mathcal{N}}\dfrac{\sum\limits_{k \in \mathcal{K}}\big(B_{S,n}^O(B_{S,k,n}^I)+\bar{B}_{k,n}^O\big)}{\big(\lambda_1\lVert\boldsymbol{v}_n^U\rVert^3+\frac{\lambda_2}{\Vert\boldsymbol{v}_n^U\rVert}\big(1+\frac{\lVert\boldsymbol{a}_n^U\rVert^2}{g^2}\big)\big)}\\[2pt]
&&\hspace{-1.2cm}\text{s.t.}\hspace{+0.4cm} z_{k,n}=\{0,1\}, \\[2pt]
&& \hspace{-0.4cm}\begin{aligned}
&\frac{\bar{B}_{k,n}^I}{\bar{R}_{k,n}^{ul}(\boldsymbol{q}_{n}^{U},z_{k,n})}+\frac{\bar{V}_{k,n}}{f_k(z_{k,n}F_L+(1-z_{k,n})F_U)}\\&+\frac{V_{S,n}}{f_SF_L}+\frac{\bar{B}_{k,n}^O}{\bar{R}_{k,n}^{dl}(\boldsymbol{q}_{n}^{U},z_{k,n})} \leq \Delta-T_{S,n}^{ul}-T_{S,n}^{dl}, \end{aligned}\\[2pt]
&&\hspace{-0.4cm} (1-c_k)\frac{B_{S,k,n}^I}{R_{S,k,n}^{ul}} \leq T_{S,n}^{ul}, \\[2pt] 
&& \hspace{-0.4cm}\frac{B_{S,n}^O(B_{S,k,n}^I)}{R_{S,n}^{dl}} \leq T_{S,n}^{dl},   \\[2pt]
&& \hspace{-0.4cm}\bar{B}_{k,n}^I, {B}_{S,k,n}^I \geq {B}_{\min}^I,\\[2pt]
&&\hspace{-0.4cm}O_{S,k}\sum_{n=1}^{n_0}B_{S,k,n}^I\hspace{-4pt}=\hspace{-4pt}\sum_{n=1}^{n_0}B_{S,n}^O(B_{S,k,n}^I), \allowdisplaybreaks\\[2pt] 
&&\hspace{-0.4cm} \bar{O}_k\sum_{n=1}^{n_0}\bar{B}_{k,n}^I=\sum_{n=1}^{n_0}\bar{B}_{k,n}^O, \\[2pt]
&& \hspace{-0.4cm}B_{S,n}^I, B_{S,n}^O, \bar{B}_{k,n}^I, \bar{B}_{k,n}^O, T^{ul}_{S,n}, T^{dl}_{S,n} \geq 0, \\[2pt] 
&& \hspace{-0.4cm}\boldsymbol{q}_{1}^U=\boldsymbol{q}_I^U, \boldsymbol{q}_{N+1}^U=\boldsymbol{q}_F^U,\\[2pt]
&& \hspace{-0.4cm}
\textrm{(\ref{vel}) - (\ref{acc_max})},
\end{eqnarray}
\end{subequations}
for $\forall k \in \mathcal{K}, \forall n \in \mathcal{N}$ and $\forall n_0 \in \mathcal{N}$, where (\ref{p1}b) is the binary variable constraint pertaining to the offloading decision targeting UAV or LEO; (\ref{p1}c)-(\ref{p1}e) enforce that the execution time of the offloaded application needs to be less than or equal to the length of the time frame of $\Delta$ seconds; (\ref{p1}f) guarantees the transmission of the minimum private and shared input bits; (\ref{p1}g) and (\ref{p1}h) enforce the completion of computing of shared data and private data in each frame; (\ref{p1}i) represents the nonnegative constraints for the bits and the latency; and 
(\ref{p1}j) and (\ref{p1}k) represent the constraints for the UAV's initial and final position and the maximum velocity and acceleration, respectively. 

The optimization problem (\ref{p1}) is non-convex due to the non-convexity of the objective function (\ref{p1}a) and the constraints (\ref{p1}b) and (\ref{p1}c), the latter of which involve the offloading decision binary variable and is coupled with the multiple variables. In addition, the problem (P1) is non-linear fractional programming (FP) due to the energy efficiency (\ref{p1}a). In general, there is no standard method that solves this kind of non-convex optimization problem all at once. Therefore, we propose the AO-based method that effectively obtains the local optimal solutions, whose details are provided as belows.
\subsection{Optimization of Cache-assisted SAGIN-MEC}
In this section, the proposed algorithm is described to obtain the locally optimal solution of the problem (\ref{p1}). To address the non-convexity of (\ref{p1}), we apply AO algorithm \cite{Zhang2018AO} along with Dinkelbach and SCA techniques \cite{Jeong2018MEC, RuiEE}. Specifically, for the given shared data information $B_{S,k,n}^I$, shared data transmission latency $\boldsymbol{T}_n$ and offloading decision $z_{k,n}$, the UAV trajectory variables $\boldsymbol{q}_n^U, \boldsymbol{v}_n^U$ and $\boldsymbol{a}_n^U$ are optimized by the Dinkelbach method based on FP in the inner loop of the SCA-based algorithm. For the given UAV trajectory $\boldsymbol{q}_n^U, \boldsymbol{v}_n^U, \boldsymbol{a}_n^U$ and offloading decision $z_{k,n}$, we optimize the shared data input bit $B_{S,k,n}^I$ and shared data transmission latency $\boldsymbol{T}_n$ by solving linear programming (LP). Lastly, for the given $B_{S,k,n}^I, \boldsymbol{T}_n$ and $\{\boldsymbol{q}_n^U, \boldsymbol{v}_n^U, \boldsymbol{a}_n^U\}$, the offloading decision $z_{k,n}$ is optimized based on binary variable relaxation.
\subsubsection{Optimization of UAV Trajectory $\{\boldsymbol{q}_n^U, \boldsymbol{v}_n^U, \boldsymbol{a}_n^U\}$}
For the given shared data information and offloading decision $\{B_{S,k,n}^I, \boldsymbol{T}_n, z_{k,n}\}$, the UAV trajectory optimization of problem (P1) can be rewritten as
\begin{subequations}\label{p2}
\begin{eqnarray}
&&\hspace{-1.2cm}\textrm{(P2)}: \max_{\substack{\boldsymbol{q}_n^U,\boldsymbol{v}_n^U,\boldsymbol{a}_n^U}} \sum_{n \in \mathcal{N}}\dfrac{\sum\limits_{k \in \mathcal{K}}\big(B_{S,n}^O(B_{S,k,n}^I)+\bar{B}_{k,n}^O\big)}{\big(\lambda_1\lVert\boldsymbol{v}_n^U\rVert^3+\frac{\lambda_2}{\Vert\boldsymbol{v}_n^U\rVert}\big(1+\frac{\lVert\boldsymbol{a}_n^U\rVert^2}{g^2}\big)\big)}\\[2pt]
&& \textrm{(\ref{p1}\textrm{c}), (\ref{p1}\textrm{j}) and (\ref{p1}\textrm{k})}.
\end{eqnarray}
\end{subequations}
The objective function (\ref{p2}a) and the constraint (\ref{p1}c) are non-convex since the denominator of the objective function is non-convex with the coupled optimization variables, and the logarithm terms of $\bar{R}_{k,n}^{ul}(\boldsymbol{q}_{n}^{U},z_{k,n})$ and $\bar{R}_{k,n}^{dl}(\boldsymbol{q}_{n}^{U},z_{k,n})$ exist. To this end, we first introduce the slack variables $\{\omega_n, \tilde{R}^{ul}_{k,n}, \tilde{R}^{dl}_{k,n}\}$. We have
$\lVert \boldsymbol{v}_n^U \rVert^2 \geq \omega_n^2, \forall n$, where $\omega_n$ is constrained not to obtain larger objective value due to velocity variable \cite{RuiEE}. The remaining slack variables $\{\tilde{R}^{ul}_{k,n}, \tilde{R}^{dl}_{k,n}\}$ are introduced to exist in the feasible set of (\ref{p1}c) with satisfying
\begin{equation}
\label{uplink_slack}
\begin{aligned}
    &\tilde{R}^{ul}_{k,n} \leq \bar{R}_{k,n}^{ul}(\boldsymbol{q}_{n}^{U},z_{k,n})=\\&\frac{W^{ul}}{K}\log_2 \bigg( 1+\gamma_{k,n}^{ul}+\frac{A_{k,n}^{ul}}{\big(\lVert \boldsymbol{q}_n^U- \boldsymbol{q}_n^D\rVert^2+H_U^2\big)}\bigg),
\end{aligned}
\end{equation}
and
\begin{equation}
\label{downlink_slack}
\begin{aligned}
    &\tilde{R}^{dl}_{k,n} \leq \bar{R}_{k,n}^{dl}(\boldsymbol{q}_{n}^{U},z_{k,n})=\\&\frac{W^{dl}}{K}\log_2 \bigg( 1+\gamma_{k,n}^{dl}+\frac{A_{k,n}^{dl}}{\big(\lVert \boldsymbol{q}_n^U- \boldsymbol{q}_n^D\rVert^2+H_U^2\big)}\bigg),
\end{aligned}
\end{equation}
where $A_{k,n}^{ul}={p_D(1-z_{k,n})\beta_0\chi^{G2A}K}/{(N_0W^{ul})}$, $\gamma_{k,n}^{ul}={p_Dz_{k,n}h_{k,n}^{D,L}K}/{(N_0W^{ul})}$, $\gamma_{k,n}^{dl}={p_Lz_{k,n}h_{k,n}^{D,L}K}/{(N_0W^{dl})}$ and $A_{k,n}^{dl}={p_U(1-z_{k,n})\beta_0\chi^{G2A}K}/{(N_0W^{dl})}$.
Then, the problem (P2) can be reformulated as

\begin{subequations}\label{p2-1}
\begin{eqnarray}
&&\hspace{-1.2cm}\textrm{(P2-1)}: \max_{\substack{\boldsymbol{q}_n^U,\boldsymbol{v}_n^U,\boldsymbol{a}_n^U, \\ \tilde{R}^{ul}_{k,n}, \tilde{R}^{dl}_{k,n}, \omega_n}} \sum_{n \in \mathcal{N}}\dfrac{\sum\limits_{k \in \mathcal{K}}\big(B_{S,n}^O(B_{S,k,n}^I)+\bar{B}_{k,n}^O\big)}{\big(\lambda_1\lVert\boldsymbol{v}_n^U\rVert^3+\frac{\lambda_2}{\omega_n}+\frac{\lambda_2\lVert\boldsymbol{a}_n^U\rVert^2}{g^2\omega_n}\big)}\\[2pt]
&&\hspace{-1cm}\text{s.t.}\hspace{+0.6cm} \begin{aligned}
&\frac{\bar{B}_{k,n}^I}{\tilde{R}^{ul}_{k,n}}+\frac{\bar{V}_{k,n}}{f_k(z_{k,n}F_L+(1-z_{k,n})F_U)}+\frac{V_{S,n}}{f_SF_L}\\&+\frac{\bar{B}_{k,n}^O}{\tilde{R}^{dl}_{k,n}} \leq \Delta-T_{S,n}^{ul}-T_{S,n}^{dl}, \, \forall n\in \mathcal{N}, \forall k\in \mathcal{K},\end{aligned}\allowdisplaybreaks\\[2pt]
&& \lVert \boldsymbol{v}_n^U \rVert^2 \geq \omega_n^2, \quad \forall n \in \mathcal{N},\\[2pt]
&& \textrm{(\ref{p1}\textrm{j}), (\ref{p1}\textrm{k}), (\ref{uplink_slack}), (\ref{downlink_slack})}. 
\end{eqnarray}
\end{subequations}
Since $\lVert \boldsymbol{v}_n^U \rVert^2$ is convex and differentiable with respect to $\boldsymbol{v}_n^U$, we have
\begin{equation}
\label{velocity_lb}
     \lVert \boldsymbol{v}_N^U(v) \rVert^2+2(\boldsymbol{v}_n^U(v))^T(\boldsymbol{v}_n^U-\boldsymbol{v}_N^U(v)) \triangleq \Psi^{lb}_n(\boldsymbol{v}_n;\boldsymbol{v}_n(v)) \geq \omega_n^2,
\end{equation}
for any local point $\boldsymbol{v}_n^U(v)$ obtained at the $v$-th iteration of SCA algorithm. For (\ref{uplink_slack}) and (\ref{downlink_slack}), we can apply the SCA-based strategy to derive its convex approximation and the lower-bound by using the first-order Taylor expansion at the given point $\lVert \boldsymbol{q}_n^U(v)-\boldsymbol{q}_n^D\rVert^2$ in the $v$-th iteration, and we have
\begin{figure*}[!t]
\begin{align}
\label{lb_ul}
&\tilde{R}^{ul}_{k,n} \leq  \dfrac{W^{ul}}{K}\log_2 \big(1+\gamma_{k,n}^{ul}+\dfrac{A^{ul}_{k,n}}{\lVert \boldsymbol{q}_n^U(v)-\boldsymbol{q}_n^D\rVert^2+H_U^2}\big) \\
&-\dfrac{W^{ul}\big(A^{ul}_{k,n}+\gamma_{k,n}^{ul}(\lVert \boldsymbol{q}_n^U(v)-\boldsymbol{q}_n^D\rVert^2+H_U^2)\big)\big(\lVert \boldsymbol{q}_n^U-\boldsymbol{q}_n^D\rVert^2-\lVert \boldsymbol{q}_n^U(v)-\boldsymbol{q}_n^D\rVert^2\big)}{K\big((1+\gamma^{ul}_{k,n})(\lVert \boldsymbol{q}_n^U(v)-\boldsymbol{q}_n^D\rVert^2+H_U^2)+A^{ul}_{k,n}\big)\big(\lVert \boldsymbol{q}_n^U(v)-\boldsymbol{q}_n^D\rVert^2+H_U^2\big)\ln 2}\triangleq \tilde{R}^{ul,lb}_{k,n}(\boldsymbol{q}_n;\boldsymbol{q}_n(v)) \nonumber  
\end{align}
\begin{align}
\label{lb_dl}
&\tilde{R}^{dl}_{k,n} \leq  \dfrac{W^{dl}}{K}\log_2 (1+\gamma_{k,n}^{dl}+\dfrac{A^{dl}_{k,n}}{\lVert \boldsymbol{q}_n^U(v)-\boldsymbol{q}_n^D\rVert^2+H_U^2}) \\
&\dfrac{W^{ul}\big(A^{dl}_{k,n}+\gamma_{k,n}^{dl}(\lVert \boldsymbol{q}_n^U(v)-\boldsymbol{q}_n^D\rVert^2+H_U^2)\big)\big(\lVert \boldsymbol{q}_n^U-\boldsymbol{q}_n^D\rVert^2-\lVert \boldsymbol{q}_n^U(v)-\boldsymbol{q}_n^D\rVert^2\big)}{K\big((1+\gamma^{dl}_{k,n})(\lVert \boldsymbol{q}_n^U(v)-\boldsymbol{q}_n^D\rVert^2+H_U^2)+A^{dl}_{k,n}\big)\big(\lVert \boldsymbol{q}_n^U(v)-\boldsymbol{q}_n^D\rVert^2+H_U^2\big)\ln 2}\triangleq \tilde{R}^{dl,lb}_{k,n}(\boldsymbol{q}_n;\boldsymbol{q}_n(v)) \nonumber  
\end{align}
\hrulefill
\vspace{-0.5cm}
\end{figure*}
equation (\ref{lb_ul}) and (\ref{lb_dl}) on the next page.

With the $r$-th UAV location $\boldsymbol{q}_n^U(v)$ and the lower bound of (\ref{lb_ul}) and (\ref{lb_dl}), the problem (P2-1) can be reformulated


\begin{subequations}\label{p2-2}
\begin{eqnarray}
&&\hspace{-1.2cm}\textrm{(P2-2)}: \max_{\boldsymbol{z}} \sum_{n \in \mathcal{N}}\dfrac{\sum\limits_{k \in \mathcal{K}}\big(B_{S,n}^O(B_{S,k,n}^I)+\bar{B}_{k,n}^O\big)}{\big(\lambda_1\lVert\boldsymbol{v}_n^U\rVert^3+\frac{\lambda_2}{\omega_n}+\frac{\lambda_2\lVert\boldsymbol{a}_n^U\rVert^2}{g^2\omega_n}\big)}\\[2pt]
&&\hspace{-1cm}\text{s.t.}\hspace{+0.4cm} \begin{aligned}
&\frac{\bar{B}_{k,n}^I}{\tilde{R}^{ul}_{k,n}}+\frac{\bar{V}_{k,n}}{f_k(z_{k,n}F_L+(1-z_{k,n})F_U)}+\frac{V_{S,n}}{f_SF_L}\\&+\frac{\bar{B}_{k,n}^O}{\tilde{R}^{dl}_{k,n}} \leq \Delta-T_{S,n}^{ul}-T_{S,n}^{dl}, \forall n\in \mathcal{N}, \forall k\in \mathcal{K},\end{aligned}\\[2pt]
&& \textrm{(\ref{p1}j), (\ref{p1}k), (\ref{velocity_lb})}, (\ref{lb_ul}), (\ref{lb_dl}),
\end{eqnarray}
\end{subequations}
where $\boldsymbol{z}(v)=(\omega_n(v), \tilde{R}_{k,n}^{ul}(v), \tilde{R}_{k,n}^{dl}(v), \boldsymbol{q}_n^U(v), \boldsymbol{v}_n^U(v), \boldsymbol{a}_n^U(v))\\\in \mathcal{X}$ for the $v$-th iterate within the feasible set of (P2-2). 
The problem (P2-2) is still non-convex owing to the objective function with the fractional form. To this end, we can adopt Dinkelbach algorithm \cite{dinkelbach} \cite{FP_SCA}, by which the objective function (\ref{p2-2}a) can be rewritten as 
\begin{equation}
    \begin{aligned}
    &F^{v}(\boldsymbol{\alpha})=\max_{\boldsymbol{z}}\bigg\{ \sum_{n \in \mathcal{N}} \bigg(2\alpha_n\sqrt{\sum_{k \in \mathcal{K}}(B_{S,n}^O(B_{S,k,n}^I)+\bar{B}_{k,n}^O)}\\&-\alpha_n^2\big(\lambda_1\lVert\boldsymbol{v}_n^U\rVert^3+\frac{\lambda_2}{\omega_n}+\frac{\lambda_2\lVert\boldsymbol{a}_n^U\rVert^2}{g^2\omega_n}\big)\bigg)|\boldsymbol{z}\in \mathcal{F}^v\bigg\},
    \end{aligned}
\end{equation}
where $\boldsymbol{z}=\{\omega_n, \tilde{R}_{k,n}^{ul}, \tilde{R}_{k,n}^{dl}, \boldsymbol{q}_n^U, \boldsymbol{v}_n^U,\boldsymbol{a}_n^U\}$, $\boldsymbol{\alpha}$ is a collection of variables $\{\alpha_1,\cdots,\alpha_N\}$, and $\mathcal{F}^v(\alpha)$ is the feasible set of problem (P2-2) at the $v$-th iteration, and is a monotonic increasing function of $\boldsymbol{\alpha}$. By \cite{FP}, the optimal ${\alpha_n}$ can be found in the closed-form as
\begin{equation}
\label{dinkel}
    {\alpha}_n^*=\frac{\sqrt{\sum_{k \in \mathcal{K}}(B_{S,n}^O(B_{S,k,n}^I)+\bar{B}_{k,n}^O)}}{\lambda_1\lVert\boldsymbol{v}_n^U\rVert^3+\frac{\lambda_2}{\omega_n}+\frac{\lambda_2\lVert\boldsymbol{a}_n^U\rVert^2}{g^2\omega_n}}, \quad \forall n \in \mathcal{N}.
\end{equation}
Based on the following Lemma 1, the iteration time can be further reduced \cite{FP_SCA} for using both SCA method and Dinkelbach algorithm.\\
\textit{\underline{Lemma} 1:} When denoting the optimal Dinkelbach parameter $\boldsymbol{\alpha}^*$ for SCA iterations by $\boldsymbol{\alpha}(v-1)$ and $\boldsymbol{\alpha}^*(v)$, we have $\boldsymbol{\alpha}^*(v-1) \leq \boldsymbol{\alpha}^*(v)$, and $F^v(\boldsymbol{\alpha}^*(v-1)) \geq F^v(\boldsymbol{\alpha}^*(v)) = 0$.\\
\textit{Proof}: See Appendix.\\
In Algorithm 1, the SCA-based process for optimizing the UAV trajectory of (P2) is described. First, the slack variables for the proposed Algorithm 1 are initialized. Next, for the $(v-1)$-th iteration, the unique Dinkelbach parameter $\boldsymbol{\alpha}^*$ is derived from (\ref{dinkel}), and (P2-2) can be solved by the conventional solver, e.g., CVX \cite{cvx}. This procedure is repeated until the optimization variables approach to the stationary point.

\begin{algorithm}[t]
	\caption{Algorithm for Optimizing the UAV's trajectory (P2-2).}
	\textbf{Input:} Introduce $\omega_n, \tilde{R}_{k,n}^{ul}, \tilde{R}_{k,n}^{dl}$, $\boldsymbol{z}(0)=\{\boldsymbol{z}_n(0)\}_{n\in \mathcal{N}} \in \mathcal{X}$ 
 
 with $\boldsymbol{z}_n(0) \triangleq ({\omega_n(0), \tilde{R}_{k,n}^{ul}(0), \tilde{R}_{k,n}^{dl}(0), \boldsymbol{q}_n^U(0), \boldsymbol{v}_n^U(0),\boldsymbol{a}_n^U(0)})$.
 
 Set $v=0$.\\
        \textbf{Output:} ${\omega_n, \tilde{R}_{k,n}^{ul}, \tilde{R}_{k,n}^{dl},\boldsymbol{q}_n^U, \boldsymbol{v}_n^U,\boldsymbol{a}_n^U}$.\\
        \textbf{Initialize:} $\boldsymbol{\alpha}^0=0$.
	\begin{algorithmic}[1]
        \Repeat
            \State $u=0$, $\boldsymbol{\alpha}^0=\boldsymbol{\alpha}^*$ in loop $v-1$;
            \Repeat
                \State Update the Dinkelbach auxiliary variable $\boldsymbol{\alpha}^*$ using 
                
                \quad (\ref{dinkel});
                \State Update $\boldsymbol{z}$ by solving the (P2-2) with (23) for the
                
                \quad fixed $\boldsymbol{\alpha}$.
                \State $u \leftarrow u+1$.
            \Until $F^v(\boldsymbol{\alpha})\leq \theta_2$
            \State $v \leftarrow v+1$.
        \Until $\boldsymbol{z}(v)$ is a stationary point of (P2-2)
        \end{algorithmic} 
\end{algorithm}
\subsubsection{Optimization of Shared Data Bits $\{B_{S,k,n}^I\}$ and latency $\{\boldsymbol{T}_n\}$}
For any given offloading decision and UAV trajectory $\{z_{k,n}, \boldsymbol{q}_n^U, \boldsymbol{v}_n^U, \boldsymbol{a}_n^U\}$, the shared data input bit $B_{S,k,n}^I$ and the transmission latency $\boldsymbol{T}_n$ can be optimized by using
\begin{subequations}\label{p3}
\begin{eqnarray}
&&\hspace{-1.5cm}\textrm{(P3)}: \max_{\substack{B_{S,k,n}^I,\boldsymbol{T}_n}} \sum_{n \in \mathcal{N}}\dfrac{\sum\limits_{k \in \mathcal{K}}\big(B_{S,n}^O(B_{S,k,n}^I)+\bar{B}_{k,n}^O\big)}{\big(\lambda_1\lVert\boldsymbol{v}_n^U\rVert^3+\frac{\lambda_2}{\Vert\boldsymbol{v}_n^U\rVert}\big(1+\frac{\lVert\boldsymbol{a}_n^U\rVert^2}{g^2}\big)\big)}\\[2pt]
&& \textrm{(\ref{p1}\textrm{c}) - (\ref{p1}\textrm{i})}.
\end{eqnarray}
\end{subequations}
Since both the objective function and constraints of (P3) are in the form of the linear combination of shared input data bit $B_{S,k,n}^I$ and delay constraint $\boldsymbol{T}_n$, the problem (P3) is a standard LP, and can be solved efficiently by the optimization tools such as CVX \cite{cvx}.
\subsubsection{Optimization of Offloading Decision $\{z_{k,n}\}$}
It is obvious that the optimization problem of offloading decisions is a mixed integer linear programming (MILP) problem. To achieve its feasible solution with low complexity, we first relax the integer constraint for the offloading decision in (\ref{p1}b). For the given shared data information and UAV trajectory $\{B_{S,k,n}^I, \boldsymbol{T}_n, \boldsymbol{q}_n^U, \boldsymbol{v}_n^U, \boldsymbol{a}_n^U\}$, the offloading decision of problem (P1) can be written as
\begin{subequations}\label{p4}
\begin{eqnarray}
&&\hspace{-1.2cm}\textrm{(P4)}: \max_{\substack{z_{k,n}}} \sum_{n \in \mathcal{N}}\dfrac{\sum\limits_{k \in \mathcal{K}}\big(B_{S,n}^O(B_{S,k,n}^I)+\bar{B}_{k,n}^O\big)}{\big(\lambda_1\lVert\boldsymbol{v}_n^U\rVert^3+\frac{\lambda_2}{\Vert\boldsymbol{v}_n^U\rVert}\big(1+\frac{\lVert\boldsymbol{a}_n^U\rVert^2}{g^2}\big)\big)}\\[2pt]
&& 0 \leq z_{k,n} \leq 1, \quad n \in \mathcal{N}, k \in \mathcal{K}, \\[2pt]
&& \textrm{(\ref{p1}\textrm{c})}.
\end{eqnarray}
\end{subequations}
In (P4), $\bar{R}_{k,n}^{ul}(\boldsymbol{q}_{n}^{U},z_{k,n})$ and $\bar{R}_{k,n}^{dl}(\boldsymbol{q}_{n}^{U},z_{k,n})$ in (\ref{p1}c) are either concave or convex with respect to the offloading decision $z_{k,n}$. To address this difficulty, we introduce the following lemma.

\textit{\underline{Lemma} 2:} Given $\gamma, B, C_1, C_2 \geq 0$ to satisfy the condition of $C_1<C_2$, the function $\psi(x) \triangleq {\gamma}/{\log_2\big(1+B(C_1-C_2)x+BC_2\big)}$ is convex within the range of $0\leq x \leq 1$.

\textit{Proof}: See Appendix.\\
Using Lemma 2, we can easily prove that $\bar{R}_{k,n}^{ul}(\boldsymbol{q}_{n}^{U},z_{k,n})$ and $\bar{R}_{k,n}^{dl}(\boldsymbol{q}_{n}^{U},z_{k,n})$ in
(\ref{p1}c) are convex with respect to $z_{k,n}$ for $0 \leq z_{k,n} \leq 1$. Therefore, (P4) is also standard LP, and can be solved efficiently, e.g., by using CVX \cite{cvx}. 

\begin{algorithm}[t]
	\caption{Proposed AO-based Algorithm for (P1).}
	\textbf{Initialize:} $\{z_{k,n}, B_{S,k,n}^I,\boldsymbol{q}_n^U,\boldsymbol{v}_n^U,\boldsymbol{a}_n^U,\boldsymbol{T}_n\}$ as arbitrary values that satisfy the constraints (\ref{p1}b)-(\ref{p1}k), Calculate the objective function in (\ref{p1}a) with the initialized variables and set $i \leftarrow 0$.
        \begin{algorithmic}[1]
        \Repeat
             \State Solve problem (P2-2) for given $\{z_{k,n}, B_{S,k,n}^I,\boldsymbol{T}_n\}$ \hspace*{\algorithmicindent}by Algorithm 1, and denote the optimal solution as \hspace*{\algorithmicindent}$\{\boldsymbol{q}_n^U,\boldsymbol{v}_n^U,\boldsymbol{a}_n^U\}$.
            \State Solve problem (P3) for given $\{z_{k,n}, \boldsymbol{q}_n^U,\boldsymbol{v}_n^U,\boldsymbol{a}_n^U\}$, and \hspace*{\algorithmicindent}denote the optimal solution as $\{B_{S,k,n}^I,\boldsymbol{T}_n\}$.
            \State Solve problem (P4) for given \hspace*{\algorithmicindent}$\{ B_{S,k,n}^I,\boldsymbol{q}_n^U,\boldsymbol{v}_n^U,\boldsymbol{a}_n^U,\boldsymbol{T}_n\}$, and denote the optimal \hspace*{\algorithmicindent}solution as $\{z_{k,n}\}$.
            \State update $i \leftarrow i+1$.
        \Until The fractional increase of the objective value is below a threshold $\epsilon>0$.
        \end{algorithmic} 
\end{algorithm}
Based on the above descriptions, we propose the AO-based Algorithm 2 to solve (P1) by alternately optimizing UAV trajectory, resource allocation and offloading decision.

\subsection{Convergence and Complexity Analysis}
\subsubsection{Convergence Analysis}
The overall Algorithm 2 is an AO-based method, and in order to guarantee its convergence, it needs to be optimally and accurately solved at each iteration in the process of updating variables for each sub-problem \cite{AO_converge}. The problems (P3) and (P4) for steps 3 and 4 of Algorithm 2 are LP-based optimization updates, guaranteeing convergence as they are optimally solved while ensuring increasing the energy efficiency \cite{Zhang2018AO}. In the case of the SCA algorithm in (P2-2), the convergence is guaranteed because the problem is feasible and the initial values of the approximation variables exist in the feasible set of the problem (P2). By following \cite{Scutari}, if the step size $\rho(v)$ is selected that satisfies $\rho \in (0,1]$, $\rho(v) \rightarrow 0$ and $\sum_{v}\rho(v)=\infty$, the optimization variable in feasible set $\boldsymbol{z}(v)$ is bounded, and the limit point $\boldsymbol{z}(\infty)$ is a stationary solution of the problem (P2-2). Additionally, the Dinkelbach algorithm can achieve the optimal $\boldsymbol{\alpha}^*$ with the linear rate \cite{FP_SCA}. It is guaranteed that the objective value of (P1) is non-decreasing over the iterations by Lemma 1. Therefore, the proposed Algorithm 2 can converge to the locally optimal solution of the problem (P1).
\subsubsection{Computational Complexity Analysis}
For each iteration of Algorithm 2, (P3) optimizes the shared input data information with the interior-point method with convex solver, and its computational complexity can be represented by $\mathcal{O}(Q_1(KN)^{3.5})$, where $Q_1$ denotes the number of iterations required to update the shared data input information. In the case of optimizing the offloading decision (P4), the complexity of solving the problem is $\mathcal{O}(Q_2(KN)^{3.5})$, where $Q_2$ denotes the number of iterations required to update the offloading decision. For the optimization of UAV trajectory in (P2), Algorithm 1 runs for the iterations of $Q_3 \times Q_4$, where the SCA algorithm's loop repeats $Q_3$ times, and the loop for the Dinkelbach algorithm repeats $Q_4$ times. Since the optimization variables handled by (P2-2) are calculated in the form of adding the slack variables to existing variables, there are total $(2KN+4N)$ variables. Therefore, the total complexity of solving the original problem (P1) is $\mathcal{O}(Q_3Q_4(2KN+4N)^{3.5})$.

\section{Numerical Results}
\begin{table}[t]
    \vspace{-2pt}
    \caption{Simulation parameter setting}
    \label{parameter}
    \centering
    \vspace{-4pt}
    \begin{tabular}{cccccc} 
        \\[-1.8ex]\hline 
        \hline \\[-1.8ex] 
        \multicolumn{1}{c}{Parameter} & \multicolumn{1}{c}{Value} & \multicolumn{1}{c}{Parameter} & \multicolumn{1}{c}{Value}\\
        \hline \\[-1.8ex] 
        {$N$} & $60$   &{$p_{D}$} &  $0.2\;$W\\
        {$\Delta$} & $7\;$s  &{$p_{U}$} &  $0.2\;$W\\
        {$K$} & $8$  &{$p_{L}$} &  $0.2\;$W\\
        {$v_L$} & $7.5\;$km/s  & $v_{\max}$ &  $50\;$m/s \\
        {$H_L$} &  $600\;$km &{$H_U$} &  $1\;$km   \\
        {$W^{ul}$} &  $40\;$MHz   &{$N_0$} &  $-174\;$dBm/Hz\\
        {$W^{dl}$} &  $40\;$MHz    &{$\beta_0$} &  $10^{-5}$    \\ 
        {$G$} &  $7000$ &{$\beta_1$} &  $10^{-5}$    \\ 
        {$f_S$} &  $1$ &{$F_L$} &  $10^{11}\,$CPU cycles/s   \\ 
        {$f_k$} &  $0.8$  &  {$F_U$} &  $5\times 10^{10}\,$CPU cycles/s     \\ 
        {$\chi^{\textrm{G2A}}$} &  $12\;$dB \cite{rician_12} & {$\varepsilon_S$} &  $2640$  \\ 
        {$\chi^{\textrm{A2A}}$} &  $30\;$dB \cite{rician_30} & {$\bar{\varepsilon}$} & $2640 \times 0.3\;$\\
        {$O_{S,k}$} &  $0.5$ & {$\bar{O}_k$} & $0.5$\\
        {$\lambda_1$} &  $9.26 \times 10^{-4}$ \cite{RuiEE} & {$\lambda_2$} & $2250$ \cite{RuiEE}\\
        {$\phi$} &  $15.8^\circ$ & ${B}_{\min}^I$ & {$5\,$}Mbits\\
        \\[-1.8ex]\hline 
        \hline \\[-1.8ex] 
    \end{tabular}
    \vspace{-10pt}
\end{table} 

In this section, we present the numerical results to validate the performance superiority of the proposed algorithm compared to the reference schemes. For reference, we consider the following schemes: (i) \textit{No Trajectory Optimization w/ Cache (NTO-C)}: In this scheme, the bit allocation and the offloading decision are optimized by the proposed algorithm, while adopting the cache with the fixed UAV trajectory; (ii) \textit{No Bit Optimization w/ Cache (NBO-C)}: The UAV path planning and offloading decision are optimized by the proposed algorithm, while adopting the cache with the fixed bit allocation; (iii) \textit{No Offloading Optimization w/ Cache (NOO-C)}: The UAV path planning and the bit allocation are optimized by the proposed algorithm, while adopting the cache with the fixed offloading decision; (iv) \textit{Joint Optimization w/ No Cache (JO-NC)}: All optimization variables are designed by the proposed algorithm without adopting the cache. For simplicity and consistency, we refer to the proposed scheme as \textit{Joint Optimization w/ Cache} (JO-C) as belows. For simulations, we consider the parameter setting by following \cite{sooyeob} and refer to the 3GPP NTN standard \cite{3GPP_NTN} for UAV and LEO settings, which is summarized in Table \ref{parameter}. For the caching model, we set the Zipf skewness factor $\varrho$ = 0.6, the number of XR dataset models $F = 3$, and assume that 2 files can be cached in LEO considering its limited cache storage. The UAV is assumed to fly from the initial spot (0, 5) km at an initial speed $\rVert \boldsymbol{v}_1^U\lVert$ = 16 m/s to the final spot (5, 10) km during the mission time. 

\begin{figure}[t] 
\centering
\includegraphics[width=1\columnwidth]{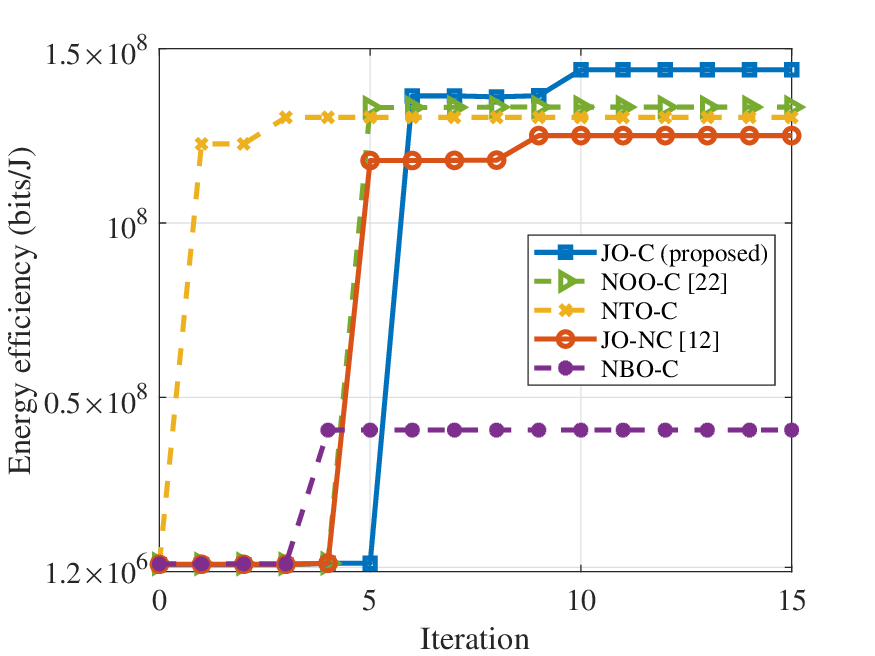}
\caption{Convergence of the proposed Algorithm 2.}
\label{iteration}
\end{figure}
\begin{figure}[t] 
\centering
\includegraphics[width=.97\columnwidth]{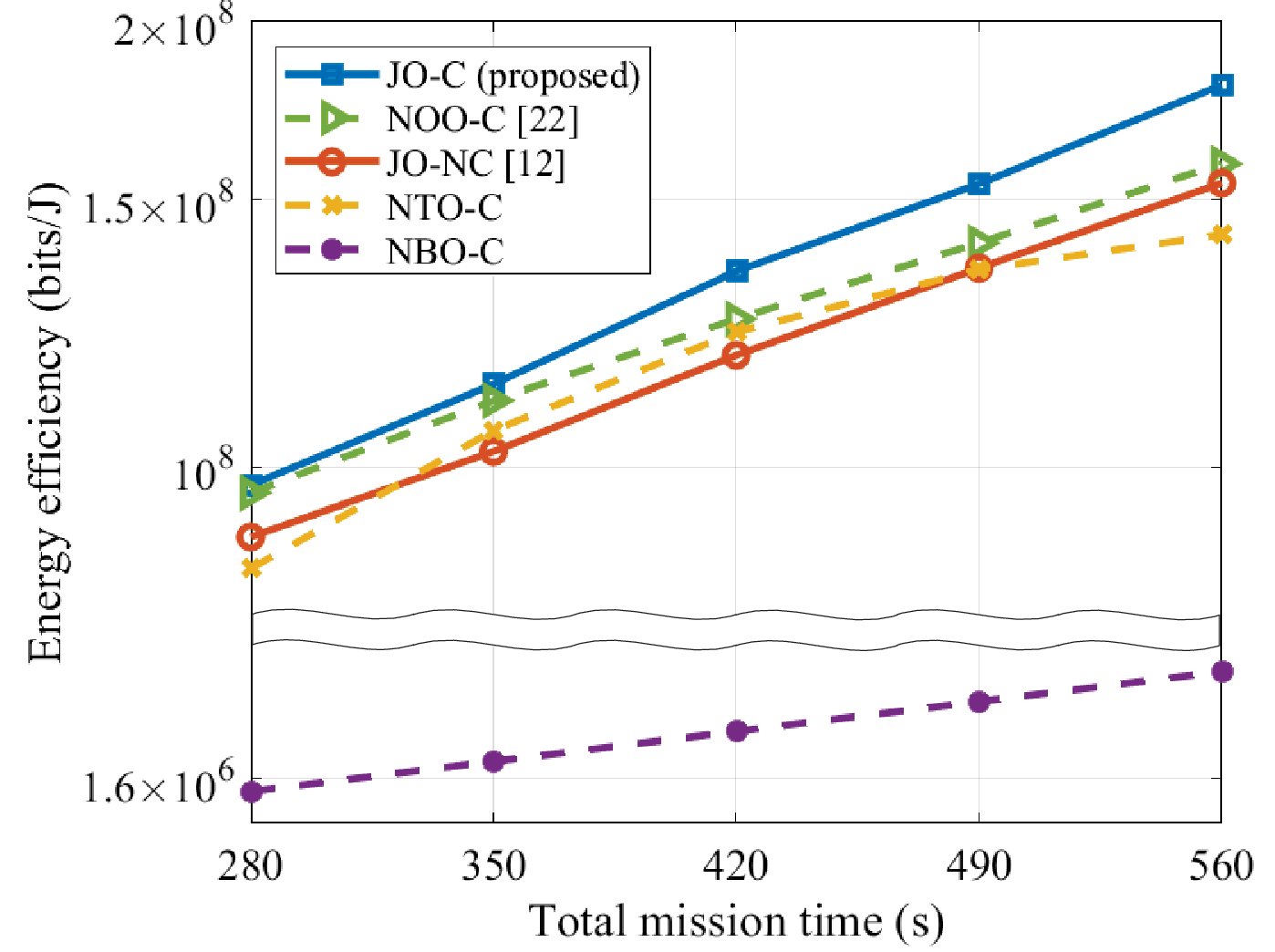}
\caption{Energy efficiency according to the different total mission time.}
\label{timeframe_sim}
\end{figure}
\subsection{Convergence and Superiority of Proposed Algorithm}
The convergence and the performance superiority of the energy efficiency of the proposed algorithm are shown with the reference schemes in Fig. \ref{iteration} and Fig. \ref{timeframe_sim}. It is observed in Fig. \ref{iteration} that all schemes converge after about 10 iterations with the AO-based algorithm. The NBO-C method, which transmits the fixed bits while satisfying the minimum bit constraints, yields the worst energy efficiency, from which the importance of shared data bit optimization can be noticed. The proposed JO-C scheme is verified to obtain the best performance by jointly optimizing all variables, which is pronounced with the larger mission time as shown in Fig. \ref{timeframe_sim}. Also, employing LEO with cache can provide the significant performance enhancement from the fact that the JO-NC scheme has the lowest energy efficiency value among all schemes except the NBO-C scheme. In Fig. \ref{timeframe_sim}, by comparing NTO-C and JO-NC schemes, the importance of cache optimization and UAV trajectory optimization is emphasized in the system design.

\begin{figure}[t] 
\centering
\includegraphics[width=1\columnwidth]{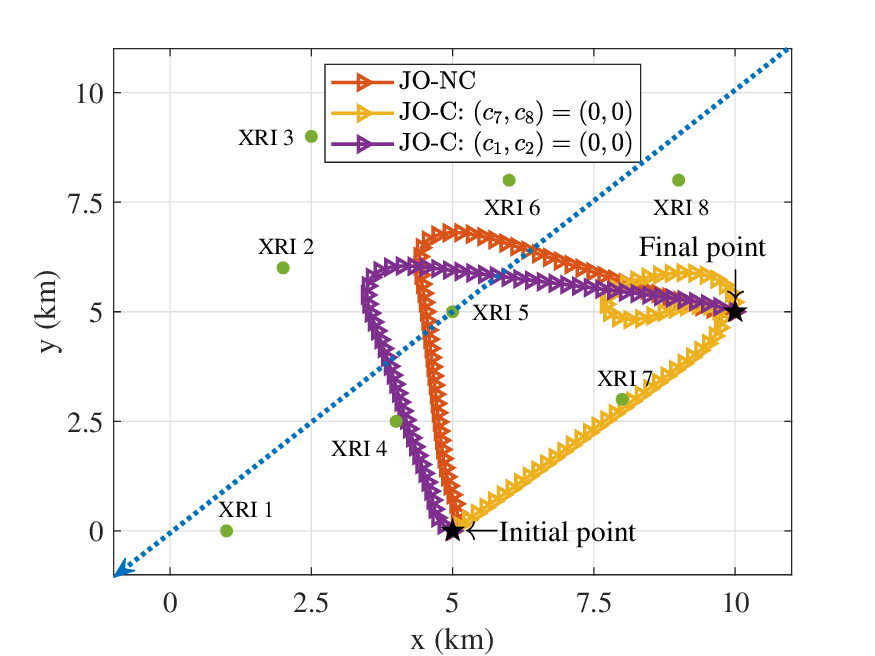}
\caption{Optimal UAV trajectories according to different caching results.}
\label{trajectory_cache}
\end{figure}

\begin{figure}[t]
\hspace{55pt}
\begin{subfigure}[t]{1\linewidth}
   \centering
   \includegraphics[width=.9\linewidth]{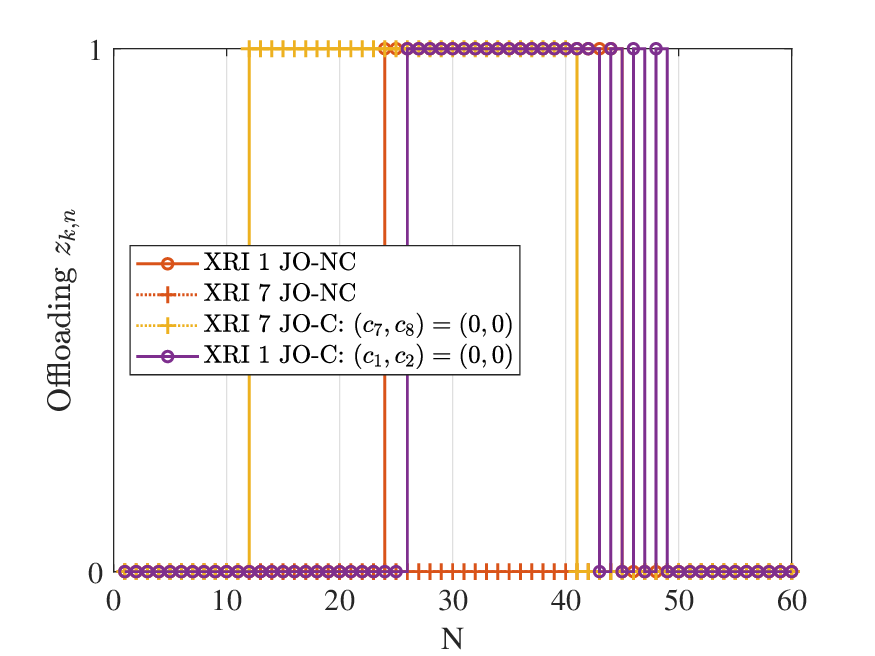}
   \caption{}
   \label{offloading}
\end{subfigure}
\vspace{-13pt}
\\[\baselineskip]
\centering
   \begin{subfigure}{1\linewidth}
   \centering
   \includegraphics[width=.9\linewidth]{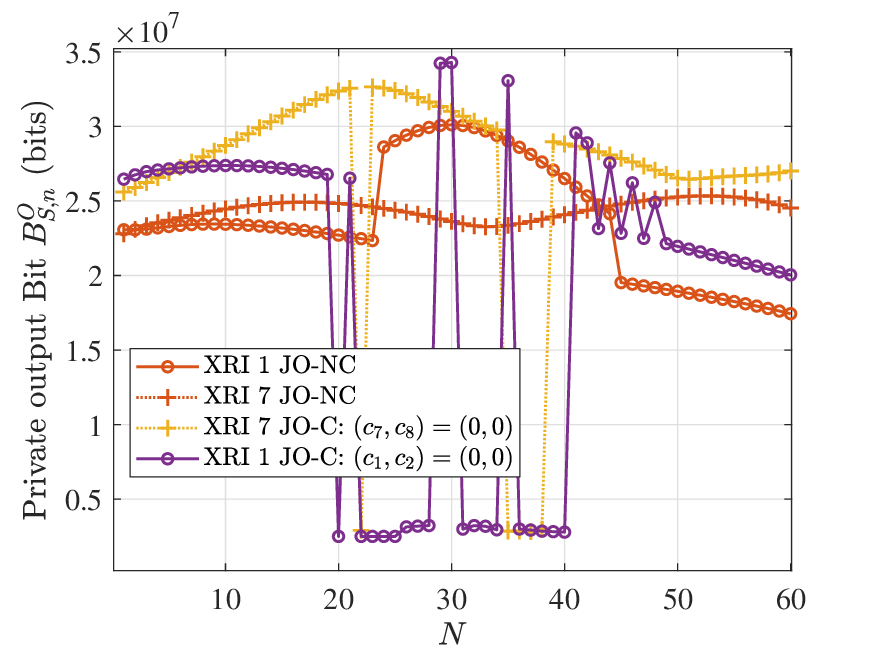}
   \caption{}
   \label{pbit} 
\end{subfigure}
\hspace{-10pt}
\begin{subfigure}{1\linewidth}
   \centering
   \includegraphics[width=.9\linewidth]{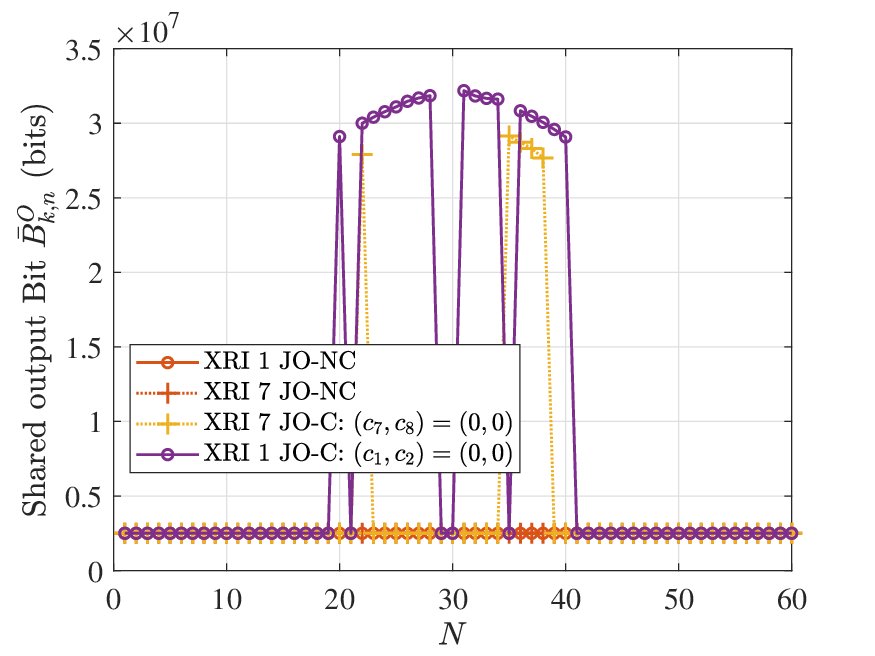}
   \caption{}
   \label{sbit}
\end{subfigure}
\centering
\caption{Optimal offloading decisions and bit allocations for XRI device 1 and 7 in Fig. \ref{trajectory_cache}.} 
\label{optimization_variable}
\end{figure}

\subsection{Impact of Caching}
Fig. \ref{trajectory_cache} shows the optimal UAV trajectories according to the different caching decision results. There are 8 XRI devices distributed randomly in a 10 km $\times$ 10 km area within the coverage of the LEO satellite. We consider three scenarios: (1) JO-NC case, where all XRI devices miss caching ($c_k$ = 0, $k \in \mathcal{K})$, (2) JO-C case with $(c_7,c_8)=(0,0)$, where, two devices rightmost from the center miss caching, i.e., $c_k = [1,1,1,1,1,1,0,0])$ and (3) JO-C case with $(c_1,c_2)=(0,0)$, where, two devices leftmost from the center miss caching, i.e., $c_k = [0,0,1,1,1,1,1,1]$. The LEO is assumed to move along the left diagonal orbit. In JO-NC scheme, the UAV moves to service the private data of all XRI devices, while minimizing the propulsion energy. On the other hand, the JO-C scheme provides the flying path staying closer to the area with the XRI devices to miss the cache, by which it consumes about 31 J more total propulsion energy than the JO-NC scheme. 

Fig. \ref{optimization_variable} illustrates the optimized offloading decision $z_{k,n}$ and output bits $B_{S,n}^O, \bar{B}_{k,n}^O$ of XRI device 1 and 7 of Fig. \ref{trajectory_cache}, where the XRI device 1 with the purple-solid line and XRI device 7 with the yellow-dashed line are represented. In Fig. \ref{optimization_variable}(a), the most of the XRI devices offload the private data to the LEO during the time frame 20 to 40, when the LEO's horizontal orbit is within the coverage area. This is because the data rate between the LEO and XRI devices increases in the middle of the mission time in order to reduce the transmission latency. Additionally, while the JO-NC scheme tends to focus on the maximization of the private data transmission by transmitting only the minimum bits $B_{\min}^I$ in shared data, the proposed JO-C scheme generates the shared output bits over $2.5 \times 10^7$ bits intermittently with transmitting the minimum private data to LEO ($z_{k,n}$ = 1) between time frame 20 and 40 as shown in Fig \ref{optimization_variable}(b) and (c). In addition, the JO-C scheme can generate more shared data output because it can allocate more downlink latency limit $T_{S,n}^{dl}$ thanks to the cache. To satisfy the latency constraint specified in (\ref{p1}c), the proposed system tends to minimize the transmission time of the private data for the cache-miss XRI devices. Accordingly, the UAV is optimized to move toward the cache-miss XRI device to maximize the private data transmission rate $\{\bar{R}_{k,n}^{ul},\bar{R}_{k,n}^{dl}\}$ as in Fig. \ref{trajectory_cache}(a). By doing so, the proposed method can maximize shared data transmission, e.g., for as the task of Mapper and Object recognizer of the XRI devices while guaranteeing the requirements for the private data transmission through caching. 

\begin{figure}[t] 
\centering
\includegraphics[width=1\columnwidth]{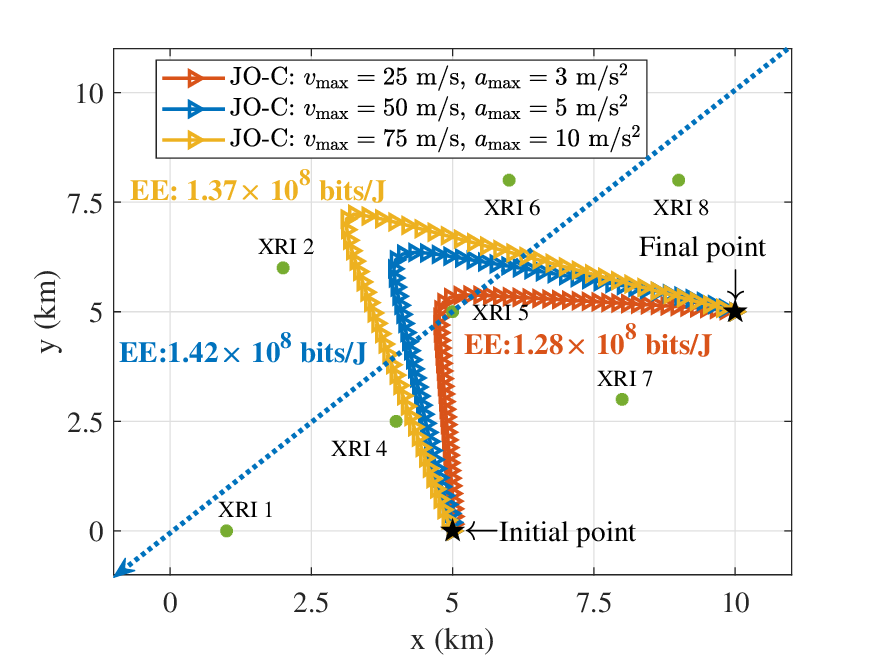}
\caption{Optimal UAV trajectories according to the different UAV maximum velocity $v_{\max}$ and maximum acceleration $a_{\max}$.}
\label{trajectory_uavvel}
\end{figure}
\begin{figure}[t] 
\centering
\includegraphics[width=1\columnwidth]{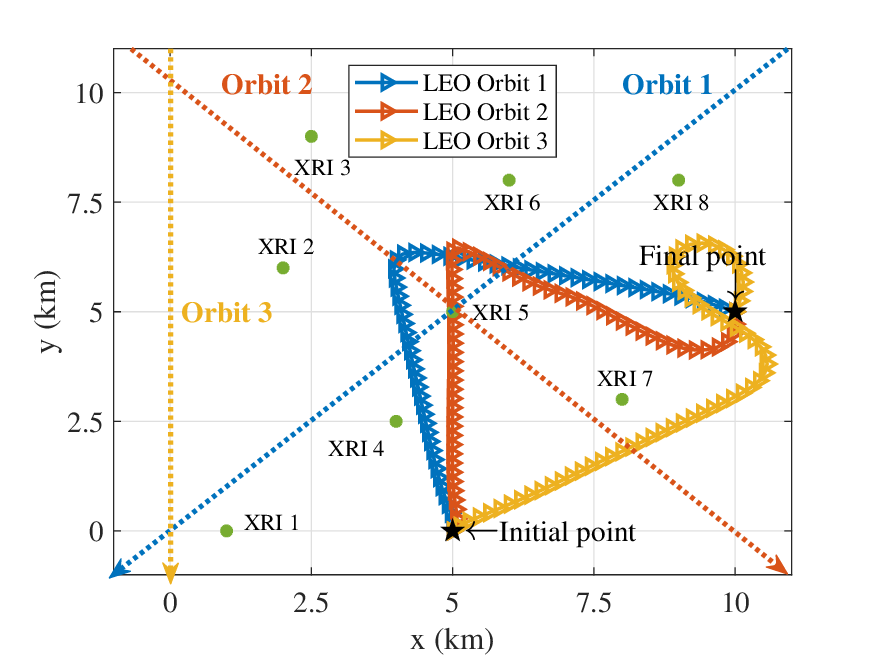}
\caption{Optimal UAV trajectories according to the different LEO satellite orbits.}
\label{trajectory_LEO}
\end{figure}
\subsection{Impact of UAV's Operation Capability}
Fig. \ref{trajectory_uavvel} illustrates the optimized UAV trajectory according to the different UAV's operationability in terms of $v_{\max}$ and $a_{\max}$. We set the caching decision scenario as $c_k$ = [1,1,1,0,0,1,1,1], where the cache-miss XRI devices are concentrated on the center. In all three cases, the UAV's acceleration $\boldsymbol{a}_n^U$ is changed significantly at the mid-frame point with the tendency to rotate and arrive at the final point. As the values of $v_{\max}$ and $a_{\max}$ increase, the point at which the flying path changes more dramatically to the upper left so as to minimize the UAV propulsion energy for satisfying the constraint (\ref{vel_max}) of (P1). In addition, the highest energy efficiency $1.42 \times 10^8$ bits/J is obtained when $v_{\max}=50$ m/s and $a_{\max}=5$ m/$\textrm{s}^2$, from which the UAV's operationabililty needs to be carefully determined for achieving the energy efficiency of SAGIN-MEC systems. 

\subsection{Impact of LEO's Orbit}
Fig. \ref{trajectory_LEO} depicts the optimized UAV trajectory according to the different LEO orbit scenarios. We set the same caching setting as Fig. \ref{trajectory_uavvel} with the 3 different LEO orbits.
In the case of the LEO Orbit 1, where the LEO moves from the upper right to the lower left of the coverage area, the UAV changes direction at coordinates (4, 6.3) km and arrives before reaching the final point. For LEO Orbit 2, the UAV's trajectory deviates slightly to the right compared to that for LEO Orbit 1. With the LEO orbit 3, where the LEO moves down along the leftmost y-axis of the map, the optimized UAV significantly moves toward the right side from the center to accommodate the XRI devices that can not be supported by LEO. This results from the fact that the optimization of the UAV trajectory $\boldsymbol{q}_n^U$ is pronounced when the UAV needs to cover the XRI devices in the area and time when LEO service is not fully available, e.g., $N<10$, $N>50$.

\section{Concluding Remarks}
In this paper, we propose the cache-assisted SAGIN-MEC for XRI applications. Particularly, based on the characteristics of XRI data, the offloaded information is assumed to be divided into shared data and private data. To mitigate redundant communication and computation of frequently occurring shared data within global coverage, we employ the caching system for the LEO satellites. The cache-miss shared data is processed by LEO, while the private data is executed by selecting either UAV or LEO considering the tradeoff between transmission and computing latency. To maximize the energy efficiency of the system, we formulate the problem to jointly optimize the UAV's trajectory, resource allocation in terms of bit and time and offloading decision, whose solution can be obtained via the AO-based method along with Dinkelbach and SCA algorithms. Via simulations, the proposed algorithm can achieve about 35\% more energy efficiency on average compared to the partial optimization schemes. In addition, use of cache can provide for 15\% performance improvement compared to the case without the cache. Future works may explore the scenario, where the multiple LEO satellites and UAVs support various XRI applications.



\appendix
\subsection{Proof of Lemma 1}
We denote the slack variable set as $\mathbb{A}(v)$. Also, we define the primary optimization variable set and corresponding Dinkelbach parameter determined in iteration $v-1$ as $\boldsymbol{z}^*(v-1)$ and $\alpha^*(v-1)$, respectively. By following \cite{dinkelbach}, $\alpha^*(v-1) 
\leq \eta(\boldsymbol{z}^*(v))$ is satisfied, where $\eta(\boldsymbol{z}^*(v))$ represents the objective function value in iteration $v$. We can obtain $\eta(\boldsymbol{z}^*(v))=\eta(\boldsymbol{z}^*(v-1);\mathbb{A}(v))$ according to the definition of $\boldsymbol{z}^*(v-1)$ and $\mathbb{A}(v)$. Since the approximated function provides the global lower
bound of the original optimization function, and the results have
to belong to the feasible set of the approximate optimization
function for the next iteration, we have $\eta(\boldsymbol{z}^*(v-1);\mathbb{A}(v)) \leq \eta(\boldsymbol{z}^*(v);\mathbb{A}(v))=\alpha^*(v)$. Therefore, $\boldsymbol{\alpha}^*(v-1) \leq \boldsymbol{\alpha}^*(v)$ holds, and it can be proved that $F^v(\boldsymbol{\alpha}^*(v-1)) \geq F^v(\boldsymbol{\alpha}^*(v)) = 0$ due to monotonically decreasing nature of $F(\alpha)$.


\subsection{Proof of Lemma 2}

We define $\xi(x) \triangleq 1+B(C_1-C_2)x+BC_2$, where $B, C_1, C_2 \geq 0$ to satisfy the condition of $C_1<C_2$, and $\psi(x)=\gamma\log_2(e)/\ln(\xi(x))$. It can be obtained that the first-order derivatives of $\psi(x)$ with respect to $x$ is
\begin{equation}
    \frac{d\psi(x)}{dx}=\frac{d\psi(x)}{d\xi(x)}\cdot\frac{d\xi(x)}{dx}=-\frac{\gamma \log_2 e\times B(C_1-C_2)}{\big(\ln(\xi(x))\big)^2\xi(x)},
\end{equation}
and the second-order derivatives of $\psi(x)$ with respect to $x$ always non-negative since
\begin{equation}
\begin{aligned}
    &\frac{d^2\psi(x)}{dx^2}=\frac{d}{dx}\bigg(\frac{d\psi(x)}{dx}\bigg)=\frac{d}{d\xi(x)}\bigg(\frac{d\psi(x)}{dx}\bigg)\cdot \frac{d\xi(x)}{dx}=\\&\gamma \log_2 e \times \big(B(C_1-C2)\big)^2 \times \frac{\big(\ln(\xi(x))\big)^2+2\ln(\xi(x))}{\big(\ln\big(\xi(x)\big)^2\xi(x)\big)^2} \geq 0,
\end{aligned}
\end{equation}
where $\ln(\xi(x))>0$ for $0 \leq x \leq 1$. Therefore, we conclude that $\psi(x)$ is a convex function.

\bibliographystyle{IEEEtran}
\bibliography{ref}

\begin{thebibliography}{10}
\providecommand{\url}[1]{#1}
\csname url@samestyle\endcsname
\providecommand{\newblock}{\relax}
\providecommand{\bibinfo}[2]{#2}
\providecommand{\BIBentrySTDinterwordspacing}{\spaceskip=0pt\relax}
\providecommand{\BIBentryALTinterwordstretchfactor}{4}
\providecommand{\BIBentryALTinterwordspacing}{\spaceskip=\fontdimen2\font plus
\BIBentryALTinterwordstretchfactor\fontdimen3\font minus \fontdimen4\font\relax}
\providecommand{\BIBforeignlanguage}[2]{{%
\expandafter\ifx\csname l@#1\endcsname\relax
\typeout{** WARNING: IEEEtran.bst: No hyphenation pattern has been}%
\typeout{** loaded for the language `#1'. Using the pattern for}%
\typeout{** the default language instead.}%
\else
\language=\csname l@#1\endcsname
\fi
#2}}
\providecommand{\BIBdecl}{\relax}
\BIBdecl

\bibitem{6G}
W.~Saad, M.~Bennis, and M.~Chen, ``A vision of 6{G} wireless systems: Applications, trends, technologies, and open research problems,'' \emph{IEEE Netw.}, vol.~34, no.~3, pp. 134--142, May 2020.

\bibitem{6G_IoT}
D.~C. Nguyen \emph{et~al.}, ``6{G} {I}nternet-of-{T}hings: A comprehensive survey,'' \emph{IEEE Internet Things J.}, vol.~9, no.~1, pp. 359--383, Jan. 2022.

\bibitem{IoT_XR_GEM}
J.~Guan and A.~Morris, ``Extended-{XRI} body interfaces for hyper-connected metaverse environments,'' in \emph{2022 IEEE Games, Entertainment, Media Conference (GEM)}, 2022, pp. 1--6.

\bibitem{IoT_XR_VRW}
J.~Guan, J.~Irizawa, and A.~Morris, ``Extended reality and {I}nternet of {T}hings for hyper-connected metaverse environments,'' in \emph{2022 IEEE Conference on Virtual Reality and 3D User Interfaces Abstracts and Workshops (VRW)}, 2022, pp. 163--168.

\bibitem{ar_enabled_IoT}
D.~Jo and G.~J. Kim, ``A{R} enabled {I}o{T} for a smart and interactive environment: A survey and future directions,'' \emph{Sensors}, vol.~19, no.~19, p. 4330, 2019.

\bibitem{IoT_XR_IoTJ}
Y.~Liu, M.~Peng, G.~Shou, Y.~Chen, and S.~Chen, ``Toward edge intelligence: Multiaccess edge computing for 5{G} and {I}nternet of {T}hings,'' \emph{IEEE Internet Things J.}, vol.~7, no.~8, pp. 6722--6747, Aug. 2020.

\bibitem{survey_5G_iot}
L.~Chettri and R.~Bera, ``A comprehensive survey on {I}nternet of {T}hings ({I}o{T}) toward 5{G} wireless systems,'' \emph{IEEE Internet Things J.}, vol.~7, no.~1, pp. 16--32, Jan. 2020.

\bibitem{iot_mec}
J.~Pan and J.~McElhannon, ``Future edge cloud and edge computing for {I}nternet of {T}hings applications,'' \emph{IEEE Internet Things J.}, vol.~5, no.~1, pp. 439--449, Feb. 2018.

\bibitem{osvaldo}
A.~Al-Shuwaili and O.~Simeone, ``Energy-efficient resource allocation for mobile edge computing-based augmented reality applications,'' \emph{IEEE Wireless Commun. Lett.}, vol.~6, no.~3, pp. 398--401, Jun. 2017.

\bibitem{globecom_xr}
X.~He, H.~Xing, Y.~Chen, and A.~Nallanathan, ``Energy-efficient mobile-edge computation offloading for applications with shared data,'' in \emph{Proc. IEEE Global Commun. Conf. (GLOBECOM)}, Abu Dhabi, United Arab Emirates, Dec. 2018, pp. 1--6.

\bibitem{xr_mec}
J.~Ren, Y.~He, G.~Huang, G.~Yu, Y.~Cai, and Z.~Zhang, ``An edge-computing based architecture for mobile augmented reality,'' \emph{IEEE Netw.}, vol.~33, no.~4, pp. 162--169, 2019.

\bibitem{powercontrol_ar}
J.~Ahn, J.~Lee, S.~Yoon, and J.~K. Choi, ``A novel resolution and power control scheme for energy-efficient mobile augmented reality applications in mobile edge computing,'' \emph{IEEE Wireless Commun. Lett.}, vol.~9, no.~6, pp. 750--754, Jun. 2020.

\bibitem{collabo_xr}
T.~Verbelen, P.~Simoens, F.~D. Turck, and B.~Dhoedt, ``Leveraging cloudlets for immersive collaborative applications,'' \emph{IEEE Pervasive Comput.}, vol.~12, no.~04, pp. 30--38, Oct./Dec 2013.

\bibitem{sooyeob}
S.~Jung, S.~Jeong, J.~Kang, and J.~Kang, ``Marine {I}o{T} systems with space-air-sea integrated networks: Hybrid {LEO} and {UAV} edge computing,'' \emph{IEEE Internet Things. J.}, pp. 1--1, Jun. 2023.

\bibitem{sagin_mec1}
S.~Mao, S.~He, and J.~Wu, ``Joint {UAV} position optimization and resource scheduling in space-air-ground integrated networks with mixed cloud-edge computing,'' \emph{IEEE Syst. J.}, vol.~15, no.~3, pp. 3992--4002, Sep. 2021.

\bibitem{ADMM}
Q.~Tang, Z.~Fei, B.~Li, and Z.~Han, ``Computation offloading in {LEO} satellite networks with hybrid cloud and edge computing,'' \emph{IEEE Internet Things J.}, vol.~8, no.~11, pp. 9164--9176, Jun. 2021.

\bibitem{sagin_mec11}
R.~Xie, Q.~Tang, Q.~Wang, X.~Liu, F.~R. Yu, and T.~Huang, ``Satellite-terrestrial integrated edge computing networks: Architecture, challenges, and open issues,'' \emph{IEEE New.}, vol.~34, no.~3, pp. 224--231, May/Jun. 2020.

\bibitem{sagin_mec2}
Z.~Zhang, W.~Zhang, and F.-H. Tseng, ``Satellite mobile edge computing: Improving {Q}o{S} of high-speed satellite-terrestrial networks using edge computing techniques,'' \emph{IEEE Netw.}, vol.~33, no.~1, pp. 70--76, Jan./Feb. 2019.

\bibitem{Jeong2018MEC}
S.~Jeong, O.~Simeone, and J.~Kang, ``Mobile edge computing via a {UAV}-mounted cloudlet: Optimization of bit allocation and path planning,'' \emph{IEEE Trans. Veh. Technol.}, vol.~67, no.~3, pp. 2049--2063, Mar. 2018.

\bibitem{sagin_company}
N.~Pachler, I.~del Portillo, E.~F. Crawley, and B.~G. Cameron, ``An updated comparison of four low earth orbit satellite constellation systems to provide global broadband,'' in \emph{Proc. IEEE Int. Conf. Commun. Workshops (ICC Workshops)}, 2021, pp. 1--7.

\bibitem{Antennagain}
Z.~Jia, M.~Sheng, J.~Li, D.~Niyato, and Z.~Han, ``{LEO}-satellite-assisted {UAV}: Joint trajectory and data collection for {I}nternet of remote things in 6{G} aerial access networks,'' \emph{IEEE Internet Things. J.}, vol.~8, no.~12, pp. 9814--9826, Jun. 2021.

\bibitem{cache_highref2}
Z.~Zhao, M.~Peng, Z.~Ding, W.~Wang, and H.~V. Poor, ``Cluster content caching: An energy-efficient approach to improve quality of service in cloud radio access networks,'' \emph{IEEE J. Sel. Areas Commun.}, vol.~34, no.~5, pp. 1207--1221, May 2016.

\bibitem{cache_highref}
E.~Bastug, M.~Bennis, and M.~Debbah, ``Living on the edge: The role of proactive caching in 5{G} wireless networks,'' \emph{IEEE Commun. Mag.}, vol.~52, no.~8, pp. 82--89, Aug. 2014.

\bibitem{satellite_cache}
D.-H. Tran, S.~Chatzinotas, and B.~Ottersten, ``Satellite- and cache-assisted {UAV}: A joint cache placement, resource allocation, and trajectory optimization for 6{G} aerial networks,'' \emph{IEEE Open J. Veh. Technol.}, vol.~3, pp. 40--54, 2022.

\bibitem{leo_cache}
C.~Qiu, H.~Yao, F.~R. Yu, F.~Xu, and C.~Zhao, ``Deep {Q}-learning aided networking, caching, and computing resources allocation in software-defined satellite-terrestrial networks,'' \emph{IEEE Trans. Veh. Technol.}, vol.~68, no.~6, pp. 5871--5883, Jun. 2019.

\bibitem{offload_cache}
Y.~Hao, M.~Chen, L.~Hu, M.~S. Hossain, and A.~Ghoneim, ``Energy efficient task caching and offloading for mobile edge computing,'' \emph{IEEE Access}, vol.~6, pp. 11\,365--11\,373, 2018.

\bibitem{vr_caching}
M.~Li, J.~Gao, C.~Zhou, X.~Shen, and W.~Zhuang, ``User dynamics-aware edge caching and computing for mobile virtual reality,'' \emph{IEEE J. Selected Topics in Signal Processing}, pp. 1--13, 2023.

\bibitem{cache_flow}
S.~Chen, L.~Rui, Z.~Gao, W.~Li, and X.~Qiu, ``Cache-assisted collaborative task offloading and resource allocation strategy: A metareinforcement learning approach,'' \emph{IEEE Internet Things J.}, vol.~9, no.~20, pp. 19\,823--19\,842, 2022.

\bibitem{vr_transcoding_cache}
H.~Xiao \emph{et~al.}, ``A transcoding-enabled 360° {VR} video caching and delivery framework for edge-enhanced next-generation wireless networks,'' \emph{IEEE J. Sel. Areas Commun.}, vol.~40, no.~5, pp. 1615--1631, May 2022.

\bibitem{chen2022joint}
J.~Chen, H.~Xing, X.~Lin, A.~Nallanathan, and S.~Bi, ``Joint resource allocation and cache placement for location-aware multi-user mobile edge computing,'' \emph{IEEE Internet Things. J.}, vol.~9, no.~24, pp. 25\,698--25\,714, 2022.

\bibitem{dinkelbach}
W.~Dinkelbach, ``On nonlinear fractional programming,'' \emph{Manage. Sci.}, vol.~13, no.~7, pp. 492--498, Mar. 1967.

\bibitem{FP_SCA}
M.~Li, N.~Cheng, J.~Gao, Y.~Wang, L.~Zhao, and X.~Shen, ``Energy-efficient {UAV}-assisted mobile edge computing: Resource allocation and trajectory optimization,'' \emph{IEEE Trans. Veh. Technol.}, vol.~69, no.~3, pp. 3424--3438, Mar. 2020.

\bibitem{RuiEE}
Y.~Zeng and R.~Zhang, ``Energy-efficient {UAV} communication with trajectory optimization,'' \emph{IEEE Trans. Wireless Commun.}, vol.~16, no.~6, pp. 3747--3760, Jun. 2017.

\bibitem{Zhang2018AO}
Q.~Wu, Y.~Zeng, and R.~Zhang, ``Joint trajectory and communication design for multi-{UAV} enabled wireless networks,'' \emph{IEEE Trans. on Wireless Commun.}, vol.~17, no.~3, pp. 2109--2121, Mar. 2018.

\bibitem{van2010survey}
D.~Van~Krevelen and R.~Poelman, ``A survey of augmented reality technologies, applications and limitations,'' \emph{Int. J. Virtual Reality}, vol.~9, no.~2, pp. 1--20, 2010.

\bibitem{Hybrid}
S.~Yoo, S.~Jeong, and J.~Kang, ``Hybrid {UAV}-enabled secure offloading via deep reinforcement learning,'' \emph{IEEE Wireless Commun. Lett.}, vol.~12, no.~6, pp. 972--976, Jun. 2023.

\bibitem{rician_12}
D.~W. Matolak and R.~Sun, ``Air–ground channel characterization for unmanned aircraft systems-part {III}: The suburban and near-urban environments,'' \emph{IEEE Trans. Veh. Technol.}, vol.~66, no.~8, pp. 6607--6618, Aug. 2017.

\bibitem{rician_eq}
C.~You and R.~Zhang, ``{3D} trajectory optimization in rician fading for {UAV-enabled} data harvesting,'' \emph{IEEE Trans. Wireless Commun.}, vol.~18, no.~6, pp. 3192--3207, Jun. 2019.

\bibitem{G2S_MIMO2016}
M.~De~Sanctis, E.~Cianca, G.~Araniti, I.~Bisio, and R.~Prasad, ``Satellite communications supporting {I}nternet of {R}emote {T}hings,'' \emph{IEEE Internet Things J.}, vol.~3, no.~1, pp. 113--123, Feb. 2016.

\bibitem{caching_decision}
S.~Nath and J.~Wu, ``Deep reinforcement learning for dynamic computation offloading and resource allocation in cache-assisted mobile edge computing systems,'' \emph{Intell. Converged Netw.}, vol.~1, no.~2, pp. 181--198, 2020.

\bibitem{FP}
K.~Shen and W.~Yu, ``Fractional programming for communication systems--{P}art {I}: Power control and beamforming,'' \emph{IEEE Trans. Signal Process.}, vol.~66, no.~10, pp. 2616--2630, May 2018.

\bibitem{cvx}
M.~Grant and S.~Boyd, ``{CVX}: Matlab software for disciplined convex programming, version 2.1,'' \url{http://cvxr.com/cvx}, Mar. 2014.

\bibitem{AO_converge}
D.~P. Bertsekas, ``Nonlinear programming,'' \emph{J. Oper. Res. Soc.}, vol.~48, no.~3, pp. 334--334, 1997.

\bibitem{Scutari}
G.~Scutari, F.~Facchinei, and L.~Lampariello, ``Parallel and distributed methods for constrained nonconvex optimization—{P}art {I}: Theory,'' \emph{IEEE Trans. Signal Process.}, vol.~65, no.~8, pp. 1929--1944, Oct. 2017.

\bibitem{rician_30}
M.~Hasib, S.~Kandeepan, W.~S.~T. Rowe, and A.~Al-Hourani, ``Direction-of-arrival ({D}o{A}) estimation performance for satellite applications in a multipath environment with rician fading and spatial correlation,'' \emph{Sensors}, vol.~23, no.~12, 2023.

\bibitem{3GPP_NTN}
X.~Lin, S.~Rommer, S.~Euler, E.~A. Yavuz, and R.~S. Karlsson, ``5{G} from space: An overview of 3{GPP} non-terrestrial networks,'' \emph{IEEE Commun. Stds, Mag.}, vol.~5, no.~4, pp. 147--153, Dec. 2021.

\end{thebibliography}
\end{document}